\pdfoutput=1

\PassOptionsToPackage{draft}{minted}
\documentclass[sigconf]{acmart}

\AtBeginDocument{%
	
  		\providecommand\BibTeX{{%
    		\normalfont B\kern-0.5em{\scshape i\kern-0.25em b}\kern-0.8em\TeX}}}

\setcopyright{none}
\copyrightyear{}
\acmYear{}
\acmDOI{}

\acmConference[]{}{}{}
\acmPrice{}
\acmISBN{}

\usepackage{tikz}
\usepackage{amsmath}
\usepackage{balance, amsfonts, graphicx, courier, textcomp, pifont, array, makecell, diagbox, xspace}

\usepackage{filecontents}
\usepackage{float}
\usepackage{algorithmic}

\usepackage{amsmath,amsopn}
\usepackage{subcaption}
\usepackage{endnotes}
\usepackage{microtype}
\usepackage{xspace}
\usepackage{graphicx}
\usepackage{fancyvrb}
\usepackage{multirow}
\usepackage{array}
\usepackage{underscore}
\usepackage{relsize}
\usepackage{pbox}
\usepackage{xstring}
\usepackage{tikz}
\usepackage[ruled,vlined]{algorithm2e}
\usepackage{fp}
\usepackage{siunitx}
\usepackage{flushend}
\usepackage{balance}
\usepackage{multicol}
\usepackage{hyperref}
\usepackage{booktabs}
\usepackage{makecell}
\usepackage{enumitem}
\usepackage{color, colortbl}

\makeatletter
\def\mdseries@tt{m}
\makeatother
\usepackage{minted}

\fvset{fontsize=\scriptsize,xleftmargin=8pt,numbers=left,numbersep=5pt}

\makeatletter
\def\PY@reset{\let\PY@it=\relax \let\PY@bf=\relax%
    \let\PY@ul=\relax \let\PY@tc=\relax%
    \let\PY@bc=\relax \let\PY@ff=\relax}
\def\PY@tok#1{\csname PY@tok@#1\endcsname}
\def\PY@toks#1+{\ifx\relax#1\empty\else%
    \PY@tok{#1}\expandafter\PY@toks\fi}
\def\PY@do#1{\PY@bc{\PY@tc{\PY@ul{%
    \PY@it{\PY@bf{\PY@ff{#1}}}}}}}
\def\PY#1#2{\PY@reset\PY@toks#1+\relax+\PY@do{#2}}

\expandafter\def\csname PY@tok@gd\endcsname{\def\PY@tc##1{\textcolor[rgb]{0.63,0.00,0.00}{##1}}}
\expandafter\def\csname PY@tok@gu\endcsname{\let\PY@bf=\textbf\def\PY@tc##1{\textcolor[rgb]{0.50,0.00,0.50}{##1}}}
\expandafter\def\csname PY@tok@gt\endcsname{\def\PY@tc##1{\textcolor[rgb]{0.00,0.27,0.87}{##1}}}
\expandafter\def\csname PY@tok@gs\endcsname{\let\PY@bf=\textbf}
\expandafter\def\csname PY@tok@gr\endcsname{\def\PY@tc##1{\textcolor[rgb]{1.00,0.00,0.00}{##1}}}
\expandafter\def\csname PY@tok@cm\endcsname{\let\PY@it=\textit\def\PY@tc##1{\textcolor[rgb]{0.25,0.50,0.50}{##1}}}
\expandafter\def\csname PY@tok@vg\endcsname{\def\PY@tc##1{\textcolor[rgb]{0.10,0.09,0.49}{##1}}}
\expandafter\def\csname PY@tok@vi\endcsname{\def\PY@tc##1{\textcolor[rgb]{0.10,0.09,0.49}{##1}}}
\expandafter\def\csname PY@tok@vm\endcsname{\def\PY@tc##1{\textcolor[rgb]{0.10,0.09,0.49}{##1}}}
\expandafter\def\csname PY@tok@mh\endcsname{\def\PY@tc##1{\textcolor[rgb]{0.40,0.40,0.40}{##1}}}
\expandafter\def\csname PY@tok@cs\endcsname{\let\PY@it=\textit\def\PY@tc##1{\textcolor[rgb]{0.25,0.50,0.50}{##1}}}
\expandafter\def\csname PY@tok@ge\endcsname{\let\PY@it=\textit}
\expandafter\def\csname PY@tok@vc\endcsname{\def\PY@tc##1{\textcolor[rgb]{0.10,0.09,0.49}{##1}}}
\expandafter\def\csname PY@tok@il\endcsname{\def\PY@tc##1{\textcolor[rgb]{0.40,0.40,0.40}{##1}}}
\expandafter\def\csname PY@tok@go\endcsname{\def\PY@tc##1{\textcolor[rgb]{0.53,0.53,0.53}{##1}}}
\expandafter\def\csname PY@tok@cp\endcsname{\def\PY@tc##1{\textcolor[rgb]{0.74,0.48,0.00}{##1}}}
\expandafter\def\csname PY@tok@gi\endcsname{\def\PY@tc##1{\textcolor[rgb]{0.00,0.63,0.00}{##1}}}
\expandafter\def\csname PY@tok@gh\endcsname{\let\PY@bf=\textbf\def\PY@tc##1{\textcolor[rgb]{0.00,0.00,0.50}{##1}}}
\expandafter\def\csname PY@tok@ni\endcsname{\let\PY@bf=\textbf\def\PY@tc##1{\textcolor[rgb]{0.60,0.60,0.60}{##1}}}
\expandafter\def\csname PY@tok@nl\endcsname{\def\PY@tc##1{\textcolor[rgb]{0.63,0.63,0.00}{##1}}}
\expandafter\def\csname PY@tok@nn\endcsname{\let\PY@bf=\textbf\def\PY@tc##1{\textcolor[rgb]{0.00,0.00,1.00}{##1}}}
\expandafter\def\csname PY@tok@no\endcsname{\def\PY@tc##1{\textcolor[rgb]{0.53,0.00,0.00}{##1}}}
\expandafter\def\csname PY@tok@na\endcsname{\def\PY@tc##1{\textcolor[rgb]{0.49,0.56,0.16}{##1}}}
\expandafter\def\csname PY@tok@nb\endcsname{\def\PY@tc##1{\textcolor[rgb]{0.00,0.50,0.00}{##1}}}
\expandafter\def\csname PY@tok@nc\endcsname{\let\PY@bf=\textbf\def\PY@tc##1{\textcolor[rgb]{0.00,0.00,1.00}{##1}}}
\expandafter\def\csname PY@tok@nd\endcsname{\def\PY@tc##1{\textcolor[rgb]{0.67,0.13,1.00}{##1}}}
\expandafter\def\csname PY@tok@ne\endcsname{\let\PY@bf=\textbf\def\PY@tc##1{\textcolor[rgb]{0.82,0.25,0.23}{##1}}}
\expandafter\def\csname PY@tok@nf\endcsname{\def\PY@tc##1{\textcolor[rgb]{0.00,0.00,1.00}{##1}}}
\expandafter\def\csname PY@tok@si\endcsname{\let\PY@bf=\textbf\def\PY@tc##1{\textcolor[rgb]{0.73,0.40,0.53}{##1}}}
\expandafter\def\csname PY@tok@s2\endcsname{\def\PY@tc##1{\textcolor[rgb]{0.73,0.13,0.13}{##1}}}
\expandafter\def\csname PY@tok@nt\endcsname{\let\PY@bf=\textbf\def\PY@tc##1{\textcolor[rgb]{0.00,0.50,0.00}{##1}}}
\expandafter\def\csname PY@tok@nv\endcsname{\def\PY@tc##1{\textcolor[rgb]{0.10,0.09,0.49}{##1}}}
\expandafter\def\csname PY@tok@s1\endcsname{\def\PY@tc##1{\textcolor[rgb]{0.73,0.13,0.13}{##1}}}
\expandafter\def\csname PY@tok@dl\endcsname{\def\PY@tc##1{\textcolor[rgb]{0.73,0.13,0.13}{##1}}}
\expandafter\def\csname PY@tok@ch\endcsname{\let\PY@it=\textit\def\PY@tc##1{\textcolor[rgb]{0.25,0.50,0.50}{##1}}}
\expandafter\def\csname PY@tok@m\endcsname{\def\PY@tc##1{\textcolor[rgb]{0.40,0.40,0.40}{##1}}}
\expandafter\def\csname PY@tok@gp\endcsname{\let\PY@bf=\textbf\def\PY@tc##1{\textcolor[rgb]{0.00,0.00,0.50}{##1}}}
\expandafter\def\csname PY@tok@sh\endcsname{\def\PY@tc##1{\textcolor[rgb]{0.73,0.13,0.13}{##1}}}
\expandafter\def\csname PY@tok@ow\endcsname{\let\PY@bf=\textbf\def\PY@tc##1{\textcolor[rgb]{0.67,0.13,1.00}{##1}}}
\expandafter\def\csname PY@tok@sx\endcsname{\def\PY@tc##1{\textcolor[rgb]{0.00,0.50,0.00}{##1}}}
\expandafter\def\csname PY@tok@bp\endcsname{\def\PY@tc##1{\textcolor[rgb]{0.00,0.50,0.00}{##1}}}
\expandafter\def\csname PY@tok@c1\endcsname{\let\PY@it=\textit\def\PY@tc##1{\textcolor[rgb]{0.25,0.50,0.50}{##1}}}
\expandafter\def\csname PY@tok@fm\endcsname{\def\PY@tc##1{\textcolor[rgb]{0.00,0.00,1.00}{##1}}}
\expandafter\def\csname PY@tok@o\endcsname{\def\PY@tc##1{\textcolor[rgb]{0.40,0.40,0.40}{##1}}}
\expandafter\def\csname PY@tok@kc\endcsname{\let\PY@bf=\textbf\def\PY@tc##1{\textcolor[rgb]{0.00,0.50,0.00}{##1}}}
\expandafter\def\csname PY@tok@c\endcsname{\let\PY@it=\textit\def\PY@tc##1{\textcolor[rgb]{0.25,0.50,0.50}{##1}}}
\expandafter\def\csname PY@tok@mf\endcsname{\def\PY@tc##1{\textcolor[rgb]{0.40,0.40,0.40}{##1}}}
\expandafter\def\csname PY@tok@err\endcsname{\def\PY@bc##1{\setlength{\fboxsep}{0pt}\fcolorbox[rgb]{1.00,0.00,0.00}{1,1,1}{\strut ##1}}}
\expandafter\def\csname PY@tok@mb\endcsname{\def\PY@tc##1{\textcolor[rgb]{0.40,0.40,0.40}{##1}}}
\expandafter\def\csname PY@tok@ss\endcsname{\def\PY@tc##1{\textcolor[rgb]{0.10,0.09,0.49}{##1}}}
\expandafter\def\csname PY@tok@sr\endcsname{\def\PY@tc##1{\textcolor[rgb]{0.73,0.40,0.53}{##1}}}
\expandafter\def\csname PY@tok@mo\endcsname{\def\PY@tc##1{\textcolor[rgb]{0.40,0.40,0.40}{##1}}}
\expandafter\def\csname PY@tok@kd\endcsname{\let\PY@bf=\textbf\def\PY@tc##1{\textcolor[rgb]{0.00,0.50,0.00}{##1}}}
\expandafter\def\csname PY@tok@mi\endcsname{\def\PY@tc##1{\textcolor[rgb]{0.40,0.40,0.40}{##1}}}
\expandafter\def\csname PY@tok@kn\endcsname{\let\PY@bf=\textbf\def\PY@tc##1{\textcolor[rgb]{0.00,0.50,0.00}{##1}}}
\expandafter\def\csname PY@tok@cpf\endcsname{\let\PY@it=\textit\def\PY@tc##1{\textcolor[rgb]{0.25,0.50,0.50}{##1}}}
\expandafter\def\csname PY@tok@kr\endcsname{\let\PY@bf=\textbf\def\PY@tc##1{\textcolor[rgb]{0.00,0.50,0.00}{##1}}}
\expandafter\def\csname PY@tok@s\endcsname{\def\PY@tc##1{\textcolor[rgb]{0.73,0.13,0.13}{##1}}}
\expandafter\def\csname PY@tok@kp\endcsname{\def\PY@tc##1{\textcolor[rgb]{0.00,0.50,0.00}{##1}}}
\expandafter\def\csname PY@tok@w\endcsname{\def\PY@tc##1{\textcolor[rgb]{0.73,0.73,0.73}{##1}}}
\expandafter\def\csname PY@tok@kt\endcsname{\def\PY@tc##1{\textcolor[rgb]{0.69,0.00,0.25}{##1}}}
\expandafter\def\csname PY@tok@sc\endcsname{\def\PY@tc##1{\textcolor[rgb]{0.73,0.13,0.13}{##1}}}
\expandafter\def\csname PY@tok@sb\endcsname{\def\PY@tc##1{\textcolor[rgb]{0.73,0.13,0.13}{##1}}}
\expandafter\def\csname PY@tok@sa\endcsname{\def\PY@tc##1{\textcolor[rgb]{0.73,0.13,0.13}{##1}}}
\expandafter\def\csname PY@tok@k\endcsname{\let\PY@bf=\textbf\def\PY@tc##1{\textcolor[rgb]{0.00,0.50,0.00}{##1}}}
\expandafter\def\csname PY@tok@se\endcsname{\let\PY@bf=\textbf\def\PY@tc##1{\textcolor[rgb]{0.73,0.40,0.13}{##1}}}
\expandafter\def\csname PY@tok@sd\endcsname{\let\PY@it=\textit\def\PY@tc##1{\textcolor[rgb]{0.73,0.13,0.13}{##1}}}

\makeatother

\def\Snospace~{\S{}}

\newcommand{\PP}[1]{
\vspace{2px}
{\bf \IfEndWith{#1}{.}{#1}{#1.}}
}

\usepackage{pifont}
\definecolor{Gray}{gray}{0.9}

\newcommand{\boxbeg}{
\vspace{2px}
\noindent\begin{tabular}{|l|}\hline
\begin{minipage}{3.2in}
\vspace{2px}
\noindent
}

\newcommand{\boxend}{
\vspace{2px}
\end{minipage}\\ \hline
\end{tabular}
\vspace{-10pt}
}

\newcommand{\BCSD}{\mbox{BinSD}\xspace}
\newcommand{\NoofBCSD}{\mbox{24}\xspace}
\newcommand{\NoofimpletedBCSD}{\mbox{15}\xspace}

\newcommand{\hookuparrow}{\mathrel{\rotatebox[origin=c]{90}{$\hookrightarrow$}}}

\usepackage{graphicx}
\usepackage{textcomp}
\usepackage{ulem}

\def\BibTeX{{\rm B\kern-.05em{\sc i\kern-.025em b}\kern-.08em
		T\kern-.1667em\lower.7ex\hbox{E}\kern-.125emX}}

\newcommand{\LPY}[1]{{#1}}

\newcommand{\tablecolor}{black}

\usepackage{etoolbox}
\usepackage{listings}
\definecolor{codegreen}{rgb}{0,0.6,0}
\definecolor{codegray}{rgb}{0.5,0.5,0.5}
\definecolor{codepurple}{rgb}{0.58,0,0.82}
\definecolor{backcolour}{rgb}{0.95,0.95,0.95}
\lstdefinestyle{mystyle}{
	backgroundcolor=\color{backcolour},   
	commentstyle=\color{codegreen},
	keywordstyle=\color{magenta},
	numberstyle=\tiny\color{codegray},
	stringstyle=\color{codepurple},
	basicstyle=\footnotesize\ttfamily,
	breakatwhitespace=false,         
	breaklines=true,                 
	captionpos=b,                    
	keepspaces=true,                 
	numbers=left,                    
	numbersep=2pt,                  
	showspaces=false,                
	showstringspaces=false,
	showtabs=false,                  
	tabsize=2,
	xleftmargin=\parindent,
}

\usepackage{url}

\usepackage{tikz}

\newcommand{\rcolor}{\rowcolor{gray!5 }}

\usepackage{float}

\begin{document}

\title{Understanding the AI-powered Binary Code Similarity Detection}

\ifdefined\DRAFT
 \pagestyle{fancyplain}
 \lhead{Rev.~\therev}
 \rhead{\thedate}
 \cfoot{\thepage\ of \pageref{LastPage}}
\fi


\author{Lirong Fu}
\authornotemark[1]
\email{fulirong007@zju.edu.cn}
\affiliation{%
   \institution{HangZhouDianZi University}
   \city{Hangzhou}
   \state{Zhejiang}
   \country{China}
}

\author{Peiyu Liu}
\email{liupeiyu@zju.edu.cn}
\affiliation{%
   \institution{Zhejiang University}
   \city{Hangzhou}
   \state{Zhejiang}
   \country{China}
}

\author{Wenlong Meng}
\email{mengwl@zju.edu.cn}
\affiliation{%
   \institution{Zhejiang University}
   \city{Hangzhou}
   \state{Zhejiang}
   \country{China}
}

\author{Kangjie Lu}
\email{kjlu@umn.edu}
\affiliation{%
   \institution{University of Minnesota}
   \city{Minneapolis}
   \state{Minnesota}
   \country{USA}
}

\author{Shize Zhou}
\email{3210100630@zju.edu.cn}
\affiliation{%
   \institution{Zhejiang University}
   \city{Hangzhou}
   \state{Zhejiang}
   \country{China}
}

\author{Xuhong Zhang}
\email{xuhongnever@gmail.com}
\affiliation{%
   \institution{Zhejiang University}
   \city{Hangzhou}
   \state{Zhejiang}
   \country{China}
}

\author{Wenzhi Chen}
\email{chenwz@zju.edu.cn}
\affiliation{%
   \institution{Zhejiang University}
   \city{Hangzhou}
   \state{Zhejiang}
   \country{China}
}

\author{Shouling Ji}
\email{sji@zju.edu.cn}
\affiliation{%
   \institution{Zhejiang University}
   \city{Hangzhou}
   \state{Zhejiang}
   \country{China}
}






\date{}

\begin{abstract}

AI-powered binary code similarity detection (\BCSD), which transforms intricate binary code comparison to the distance measure of code embedding through neural networks, has been widely applied to program analysis. 
However, due to the diversity of the adopted embedding strategies, evaluation methodologies, running environments, and/or benchmarks, it is difficult to quantitatively understand to what extent the \BCSD problem has been solved, especially in real-world applications.
Moreover, the lack of an in-depth investigation of the increasingly complex embedding neural networks and various evaluation methodologies has become the key factor hindering the development of AI-powered \BCSD.  

To fill these research gaps, in this paper, we present a systematic evaluation of state-of-the-art AI-powered \BCSD approaches by conducting a comprehensive comparison of \BCSD systems on similar function detection and two downstream applications, namely vulnerability search and license violation detection.
Building upon this evaluation, we perform the first investigation of embedding neural networks and evaluation methodologies. The experimental results yield several findings, which provide valuable insights in the \BCSD domain, including (1) despite the GNN-based \BCSD systems currently achieving the  best performance in similar function detection, there still exists considerable space for improvements; (2) the capability of AI-powered \BCSD approaches exhibits significant variation when applied to different downstream applications; 
(3) existing evaluation methodologies still need substantial adjustments. For instance, the evaluation metrics (such as the widely adopted ROC and AUC)  usually fall short of accurately representing the model performance of the practical use in real-world scenarios. 
Based on the extensive experiments and analysis, we further provide several promising future research directions. To facilitate future work in this interesting research, we will open source the entire datasets, benchmarks, and implementation details.      
\end{abstract}

\maketitle




\section{Introduction}
\label{s:intro}

\BCSD aims to measure the similarity of two binary programs without the corresponding source code. Currently, \BCSD has become a fundamental component supporting multiple applications, including vulnerability detection~\cite{pewny2014leveraging, eschweiler2016discovre, xu2017neural, gao2018vulseeker, chandramohan2016bingo, shirani2018binarm, liu2018alphadiff, massarelli2018safe, ding2019asm2vec,yang2021asteria}, license violation detection~\cite{luo2014semantics, zhang2014program},  patch analysis~\cite{zhang2018precise,jiang2020pdiff}, and malware analysis~\cite{comparetti2010identifying, jang2011bitshred,mirzazadeh2015metamorphic}. Particularly, \BCSD has become the de facto approach to decrease the severe consequences brought by supply-chain attacks~\cite{ eschweiler2016discovre, xu2017neural, gao2018vulseeker, chandramohan2016bingo, shirani2018binarm, liu2018alphadiff, massarelli2018safe, ding2019asm2vec,yang2021asteria}.  
Over the past few decades, researchers have proposed many remarkable \BCSD approaches that enable different capabilities and characteristics in the problem domain. 
Unfortunately, significant binary changes introduced by compilation configurations, instruction set architecture (ISA), and toolchains make \BCSD non-trivial. Moreover, with the rapid development of software engineering, modern software programs typically include an increasing number of lines of code and are widespread, resulting in an overwhelming number of programs with complex structures and making \BCSD harder than ever. 

Under the immense surge in AI technologies, including natural language processing (NLP)~\cite{chowdhary2020natural, manning1999foundations,hirschberg2015advances, Mikolov2013Distributed} and graph representation learning (GRL)~\cite{chen2020graph, ying2018hierarchical}, many research works~\cite{xu2017neural, gao2018vulseeker, yu2020order, fu2021focus, ding2019asm2vec, massarelli2018safe, Baldoni2018Unsupervised, ling2021multilevel,li2021palmtree,lee2017learning, peng2021could, yang2021codee,yang2021asteria,duan2020deepbindiff,redmond2018cross,zuo2019neural,kim2022improving, luovulhawk,ahn2022practical} are proposed to apply the widely used neural networks such as graph neural network (GNN)~\cite{fan2019graph}, convolutional neural network (CNN)~\cite{albawi2017understanding}, and long short-term memory (LSTM)~\cite{Greff2017LSTM} to solve \BCSD problem. 
%
Compared to traditional \BCSD approaches ~\cite{gao2008binhunt,ming2012ibinhunt,jin2012binary,ng2013expose,pewny2014leveraging,pewny2014leveraging,luo2014semantics,pewny2015cross,david2016statistical,feng2017extracting,shirani2018binarm,huang2017binsequence,lakhotia2013fast,david2017similarity} that employ various strategies such as symbolic execution, graph matching, hashing technique, etc., 
AI-powered \BCSD approaches have achieved considerable success~\cite{xu2017neural}. Nonetheless, understanding the extent to which the \BCSD problem has been resolved remains challenging due to the following reasons.

\textit{Inconsistent embedding strategies and evaluation methodologies.}  Due to the inconsistencies of the adopted embedding strategies, running environments, and/or benchmarks, it is difficult to understand the performance of existing AI-powered \BCSD approaches in a fair and quantitative way. Specifically, as shown in~\autoref{tab:BCSD-comparision}, we carefully select and investigate \NoofBCSD representative AI-powered \BCSD systems published in recent years. We find that each \BCSD tool adopts a different set of binary feature extraction methods, code embedding networks, and evaluation datasets, which impedes a fair understanding of the AI-powered \BCSD tools. 

\textit{Opaque neural network understanding}.  
Currently, the neural network models used to embed binaries in \BCSD are becoming increasingly complex and diverse. However, most neural networks are still treated as black-box function approximators~\cite{chakraborty2017interpretability}. Thus, the complex structure of neural networks and opaque data processing procedures may hinder the justification and users' trust when there is a lack of an in-depth model understanding. 

\textit{The lack of in-depth metric understanding}. 
In~\autoref{tab:BCSD-comparision},  16 out of the 24 \BCSD approaches employ receiver operating characteristics (ROC) curve or area under curve (AUC) values, two discrimination indexes used to evaluate classifiers to show the \BCSD performance. However, \BCSD is more similar to the recommendation process---given a query item, reporting the top-K similar search results from a predefined searching repository. It is unknown whether these machine learning metrics are sufficient for evaluating the performance of existing \BCSD approaches. Insufficient metrics would mislead users about the actual capabilities of \BCSD approaches. 

\textit{The lack of downstream application evaluation}.
%
We note that most \BCSD approaches do not actually examine their performance in real-world downstream applications, which may confuse users about their true capabilities. 
Indeed, part of \BCSD  works~\cite{xu2017neural, gao2018vulseeker, fu2021focus}  evaluated their effectiveness by vulnerability detection. However, such case studies may not be representative enough. After all, each downstream application has its own characteristics and requires different \BCSD abilities. Currently, the research community still lacks a systematic understanding of how well the AI-powered \BCSD tools support downstream applications. 

Without an in-depth investigation of the above problems, it is difficult to boost meaningful and effective progress in the \BCSD field.
To fill these research gaps, 
we first perform a comprehensive evaluation of the AI-powered \BCSD approaches by using the same benchmarks and evaluation metrics on the tasks of similar function detection and downstream application utilities. We set up two representative applications---vulnerability search and license violation detection, as they are the primary downstream applications that existing \BCSD approaches claim that they can support.
Then, we perform an in-depth analysis of the adopted binary embedding neural networks and evaluation methodologies to understand their strengths and limitations. 
In summary, we aim to investigate the following three aspects of the \BCSD research field:

\begin{itemize}
 \item  Evaluating how state-of-the-art AI-powered \BCSD methods perform in similar function detection and downstream applications;
 \item  Understanding the widely adopted embedding neural networks and evaluation methodologies  in \BCSD approaches; 
 \item {Exploring the potential promising future directions.}
\end{itemize}

To achieve these research goals, we design and implement a framework that includes representative state-of-the-art AI-powered \BCSD tools (function level approaches), two datasets, and evaluation benchmarks.  
With the framework, we conduct comprehensive evaluations and make insightful observations that amend or complement prior knowledge. The main findings include: 
(1) despite GNN-based BinSD approaches performing the best in similar function detection, 
their intrinsic limitation is the problem of ``embedding collision''---different binary functions happen to have similar function embeddings;
(2) the \BCSD capability varies when applied to different downstream applications. For instance, GNN-based BinSD approaches achieve better performance on vulnerability search, while CNN-based BinSD approaches are more applicable in the detection of license violations;
(3) the evaluation metrics used in many existing AI-powered \BCSD approaches are insufficient to represent the actual capability of an AI-powered \BCSD tool; 
(4) the problem of \BCSD has not been well addressed, and most existing \BCSD solutions still have considerable improvement space.
{Beyond the above insightful findings, we also provide several promising strategies to enhance \BCSD. For instance, the proposed embedding concatenation and graph alignment can highly likely alleviate the ``embedding collision'' problem. }
In summary, The key contributions of this paper are summarized as follows.
\begin{itemize}
    \item We present a systematic evaluation of the state-of-the-art AI-powered \BCSD approaches in a quantitative and fair way.
    The experimental results show that the current embedding neural networks and evaluation methodologies still need further improvements, implying that both subjectivity and bias may exist in the evaluation of prior \BCSD approaches.
    
    \item We employ two widely used downstream applications to validate the utilities of current AI-powered \BCSD tools. We observe that many \BCSD tools are applicable for license violation detection. By contrast, for vulnerability search, the accuracy of existing \BCSD approaches is still far from expectation. 
    
    \item We summarize the findings and implications of our study, which can help improve the understanding of the existing binaries embedding neural networks. Our findings can further provide guidelines for future improvements to the \BCSD problem.
\end{itemize}

We will open-source the artifacts to support the research community for further research and the enhancement of \BCSD.

\begin{table*}
    \footnotesize
    \color{\tablecolor}
	\centering
	\caption{Comparison among state-of-the-art AI-powered \BCSD approaches. MFE = manual feature extraction. AE = assembly extraction. IRE = intermediate representation extraction. ASTE = abstract syntax tree extraction.  PStructure2Vec = Parameterized Structure2Vec network. NGMN = node-graph matching network. Func = function level. BB = basic block level. INST = instruction level. P@K = precision at the top-K search result. DEV = deviation. 
	}\vspace{-6pt}
	\label{tab:BCSD-comparision}
	\setlength{\tabcolsep}{1mm} 
 	{
	\begin{tabular}{llcccccc}
		\toprule
		\multirow{-2}{*}{\textbf{\shortstack{Detector}}}   & \textbf{\shortstack{Feature \\Extraction }} & \textbf{\shortstack{Embedding\\Networks}} & \textbf{\shortstack{Approach\\Granularities}}  &\textbf{\shortstack{Evaluated \\Arches}} &\textbf{\shortstack{Bug \\ Detection}} &\textbf{\shortstack{License Violation \\ Detection}}&\textbf{\shortstack{Evaluated \\Metrics}}\\
		\midrule

 
                {VulHawk}~\cite{luovulhawk}& RoBERTa&
 GCN  & Func  &\shortstack{x86, x64, MIPS \\MIPS64, ARM, AArch64} &  \checkmark & \ding{55}& \shortstack{ROC~\cite{zou2007receiver}, AUC~\cite{zou2007receiver}\\ P@K~\cite{ranking}, R@K~\cite{ranking}}\\ 
  
\rcolor Gemini~\cite{xu2017neural} & MFE & PStructure2Vec  & Func   &x86, MIPS, ARM & \checkmark & \ding{55}&  ROC~\cite{zou2007receiver}, AUC~\cite{zou2007receiver}, P@K~\cite{ranking}\\ 

	    Gemini-skip~\cite{fu2021focus} & skip-gram
& PStructure2Vec  & Func  &x86, MIPS, ARM & \ding{55} & \ding{55} & ROC~\cite{zou2007receiver}, AUC~\cite{zou2007receiver}, P@K~\cite{ranking}\\ 
\rcolor			VulSeeker~\cite{gao2018vulseeker}& MFE & DNN &Func &\shortstack{x86, x64, MIPS \\MIPS64, ARM, AArch64}  &\checkmark & \ding{55}& \shortstack{ROC~\cite{zou2007receiver}, AUC~\cite{zou2007receiver}\\ ACC~\cite{zou2007receiver}, P@K~\cite{ranking}}\\
		VulSeeker-skip~\cite{fu2021focus}& skip-gram
& DNN &Func &\shortstack{x86, x64, MIPS \\MIPS64, ARM, AArch64}  &\ding{55} & \ding{55}& \shortstack{ROC~\cite{zou2007receiver}, AUC~\cite{zou2007receiver}\\ ACC~\cite{zou2007receiver}, P@K~\cite{ranking}}\\
\rcolor		Asm2Vec~\cite{ding2019asm2vec}& AE & PV-DM&Func & x86 &\checkmark & \ding{55}& P@K~\cite{ranking}, TPR\cite{zou2007receiver}\\
		
	SAFE~\cite{massarelli2018safe}& skip-gram
& SAN&Func  &x64, ARM &\checkmark & \ding{55} &ROC~\cite{zou2007receiver}, AUC~\cite{zou2007receiver}\\
\rcolor			UFE-mean~\cite{Baldoni2018Unsupervised}& i2v\_mean & Structure2Vec& Func& x86, ARM &\ding{55} & \ding{55}& ROC~\cite{zou2007receiver}, AUC~\cite{zou2007receiver}\\
 		UFE-attention~\cite{Baldoni2018Unsupervised}&  i2v\_attention & Structure2Vec& Func&x86, ARM &\ding{55} & \ding{55}& ROC~\cite{zou2007receiver}, AUC~\cite{zou2007receiver}\\
\rcolor			UFE-rnn~\cite{Baldoni2018Unsupervised}& i2v\_rnn & Structure2Vec& Func&x86, ARM &\ding{55} & \ding{55}& ROC~\cite{zou2007receiver}, AUC~\cite{zou2007receiver} \\
 	Focus~\cite{fu2021focus}& MFE & GTN&Func &\shortstack{x86, x64, MIPS, MIPS64,\\ ARM, AArch64, PPC, PPC64}  &\ding{55} & \ding{55}& \shortstack{ROC~\cite{zou2007receiver}\\ AUC~\cite{zou2007receiver}, P@K~\cite{ranking}}\\ 
\rcolor		Focus-skip~\cite{fu2021focus}& skip-gram
&  GTN&Func &\shortstack{x86, x64, MIPS, MIPS64,\\ ARM, AArch64, PPC, PPC64} &\checkmark & \ding{55}& \shortstack{ROC~\cite{zou2007receiver}\\ AUC~\cite{zou2007receiver}, P@K~\cite{ranking}}\\ 
 		BinaryAI-skipt~\cite{yu2020order}& skip thought &CNN, MPNN & Func &x64, ARM & \ding{55} & \ding{55} & Rank-1~\cite{Rank}, MRR~\cite{MRR}\\
\rcolor		BinaryAI-bert2~\cite{yu2020order}& BERT & CNN, MPNN & Func&x64, ARM &\ding{55} & \ding{55} & Rank-1~\cite{Rank}, MRR~\cite{MRR}\\
 		MGMN~\cite{ling2021multilevel}& MFE & SGNN, NGMN &Func& x86, MIPS, ARM &\ding{55} & \ding{55}&AUC~\cite{zou2007receiver}\\
\rcolor			{Oscar~\cite{peng2021could}} & IRE & Transformer &Func& x86 & \ding{55}&\ding{55}& R@1~\cite{ranking} \\
		{Codee~\cite{yang2021codee}} & skip-gram
& ADMM
&Func&  x64, MIPS, ARM &\checkmark &\ding{55}& \shortstack{ROC~\cite{zou2007receiver}\\ F1-score~\cite{zou2007receiver}, R@K~\cite{ranking}}\\
\rcolor		{Asteria~\cite{yang2021asteria}} &ASTE & Tree-LSTM &Func& x86, x64, ARM, PPC &\checkmark & \ding{55} & ROC~\cite{zou2007receiver}, AUC~\cite{zou2007receiver}\\ 
\rcolor		\textsc{DeepBinDiff~\cite{duan2020deepbindiff}}& CBOW
& TADW
&BB &x86 &\checkmark &\ding{55}& F1-score~\cite{zou2007receiver}, CDF~\cite{CDF}\\
		{RLZ~\cite{redmond2018cross}} & AE & CBOW
		&BB &x64, ARM&\ding{55}& \ding{55}&ROC~\cite{zou2007receiver}, AUC~\cite{zou2007receiver}\\
            \rcolor	{INNEREYE~\cite{zuo2019neural}}& skip-gram & LSTM & BB &x64, ARM&\ding{55}& \ding{55} &ROC~\cite{zou2007receiver}, AUC~\cite{zou2007receiver}\\
            XBA ~\cite{kim2022improving} & MFE & GCN  & BB   &x64, AArch64  &\checkmark& \ding{55}&Hit@K~\cite{liu2018alphadiff}\\  

           \rcolor {PALMTREE~\cite{li2021palmtree}}& AE & BERT &INST &x64&\ding{55} & \ding{55} &ACC~\cite{zou2007receiver}, AUC~\cite{zou2007receiver}, DEV~\cite{Deviation}\\
            	{Instruction2Vec~\cite{lee2017learning}}& AE & Text-CNN &INST &-&\ding{55}& \ding{55} & ACC~\cite{zou2007receiver}\\
		\bottomrule
	\end{tabular}}
\end{table*}

\section{ Background}
\label{s:preliminaries}
In this section, we first elaborate on the workflow of \BCSD approaches to understand the mainstream solutions. Then, we describe the paper selection process of our measurement.

\subsection{Workflow of AI-powered \BCSD}
Despite the AI-powered \BCSD approaches are often built on different techniques and complex neural networks, their mainstream workflow usually consists of three phases: preprocessing, code representation, and code embedding.  
\label{phaseofBCSD}

 \textbf{Preprocessing.} 
Given two binary code snippets in the first phase, the AI-powered \BCSD tools usually perform binary feature extraction. Specifically, after code disassembly, most \BCSD approaches~\cite{duan2020deepbindiff, fu2021focus, massarelli2018safe, zuo2019neural} perform instruction normalization to avoid the out-of-vocabulary (OOV) problem~\cite{young1994detecting} (a common issue in NLP). Besides, some \BCSD tools~\cite{chandramohan2016bingo, ding2019asm2vec} perform selective callee inline to decrease the CFG change brought by function inlining.
Several other \BCSD tools~\cite{xu2017neural, gao2018vulseeker} perform manual feature extraction or instruction embedding~\cite{massarelli2018safe, fu2021focus, yu2020order} to obtain binary features as the unique label of binaries.  

\textbf{Code representation.} After preprocessing, intuitively, one needs to consider the specific binary code representation
as the input of embedding neural networks. According to our empirical analysis, the AI-powered BinSD tools usually represent a binary code snippet as an abstract syntax tree (AST), control flow graph (CFG), data flow graph (DFG), basic block adjacent matrix, sequential instruction sequences, or the combinations of them.

\textbf{Code embedding.}  
Finally, based on the widely used neural networks such as GNN, CNN, and RNN/LSTM, the AI-powered \BCSD approaches convert specific code representation to low-dimensional vector---code embedding, which will be further compared to determine the similarity score of the input binary code pairs.

\subsection{Paper Selection}

\BCSD approaches are supposed to be used in large-scale downstream applications. In this situation, most researchers employ static analysis to ensure efficiency and avoid the code coverage problem in dynamic analysis. Thus, this paper aims to evaluate static AI-powered \BCSD approaches. Moreover,
depending on the specific downstream applications, the granularity of \BCSD  approaches varies, 
including instruction, basic block, function, and binary file comparison.
We mainly evaluate existing \BCSD approaches at the function level---providing a pair of binary functions, we examine whether they are semantically similar (compiled from the same source code) because most \BCSD approaches are based on function level comparison. Besides, the comparison of similar binary semantics implemented by different source code snippets and the analysis of the obfuscated binaries and malware (highly likely obfuscated programs) is out of our scope since most \BCSD approaches do not support such binary analysis. Finally, this paper aims to measure existing research works specifically designed for the \BCSD problem. The evaluation of various embedding networks~\cite{li2019graph, fan2019graph} that can be used to solve \BCSD problem after extra feature engineering is also out of our scope. We clarify that this paper focuses on the investigation of the mainstream techniques employed in \BCSD approaches instead of specific \BCSD methods. Thus, we do not aim for the evaluation of all the AI-powered \BCSD approaches but the representative state-of-the-art \BCSD approaches (as shown in~\autoref{tab:BCSD-comparision}) using various embedding strategies. 

\section{How  \BCSD approaches perform}
\label{s:eval}

This section comprehensively evaluates how state-of-the-art AI-powered \BCSD approaches perform in similar function detection and two representative downstream applications.

\textbf{Similar function detection.} 
This section uses the same dataset and various metrics to perform a fair and comprehensive accuracy and efficiency comparison. In the accuracy evaluation, given a query function, we investigate how many semantically similar functions (the ones compiled from the same source code as the query function) can be identified among a large number of functions in a repository by existing \BCSD approaches.  
Specifically, from the \NoofBCSD \BCSD  approaches listed in~\autoref{tab:BCSD-comparision}, we evaluate \NoofimpletedBCSD representative \BCSD approaches that perform function level comparison.
Oscar~\cite{peng2021could} is excluded because it requires too many computing resources (8 V100 GPUs). 
Besides, we also excluded Codee~\cite{yang2021codee}, XBA~\cite{kim2022improving}, and VulHawk~\cite{luovulhawk}, which are partially open-sourced and cannot be accurately implemented based on the available descriptions in the corresponding papers.
We have yet to receive a reply or obtain limited support from the authors after contacting them.

\textbf{Downstream applications.}  
%
Many researchers claim the proposed \BCSD can be widely applied to vulnerability search and license violation detection in IoT firmware images. This paper thus evaluates the selected \BCSD approaches in these two real-world scenarios. Considering we have comprehensively compared each \BCSD approach in similar function detection and the considerable manual effort cost in the bug and license violation confirmation process, we sample several \BCSD approaches to understand the actual performance of mainstream \BCSD approaches according to the following rules. (1) Support \BCSD analysis under ARM (the architecture of IoT devices analyzed in this evaluation). (2) Be applicable in large-scale analysis. For instance, it takes more than six months for Focus-skip to generate the function embedding for all IoT firmware. Thus, we exclude this method. (3) Present the mainstream function embedding techniques, including RNN/LSTM-based, GNN-based, and CNN-based models. (4) Perform well in the similar function detection task;  For instance, for Gemini variations, Gemini-skip performs the best. Thus, we include Gemini-skip in the downstream evaluation.
Finally, we select five \BCSD approaches to understand how state-of-the-art \BCSD approaches perform in real-world scenarios. 
\subsection{Implementation and Setup}
\label{sub:setup}

\textbf{Implementation:} Our implementation includes the following three aspects. (1) We correct part of the open-sourced \BCSD approaches that cannot be reproduced due to some logic errors or program implementation flaws. Moreover,  we reimplement part of closed-sourced \BCSD approaches such as BinaryAI-bert2~\cite{yu2020order}. The reimplemented models achieve similar AUC as those reported in the corresponding papers.
(2) To perform a comparative evaluation, this paper uses the same datasets to understand the capabilities of \BCSD systems. For AI-powered approaches, when changing the experiment datasets, the hyper-parameters of neural networks may also need to change to ensure the best performance. Thus, we adjust the hyper-parameters for the evaluated \BCSD embedding neural networks \LPY{to ensure each \BCSD approach presents the best performance when compared to other \BCSD tools. }
(3) To calculate evaluation metrics such as precision, recall, and MAP in similar function detection, we need to generate binary function embeddings. \LPY{Then, we can perform embedding comparisons.} However, most \BCSD approaches do not provide function embedding generation methods. 
Thus, we implement the function embedding generation for these \BCSD approaches.
How we improve/reimplement existing \BCSD approaches, select hyper-parameters, and generate function embedding are publicly available in artifacts~\footnote{https://anonymous.4open.science/r/AI-powered-BinSD-Analysis-9251/}.

%

\textbf{Setup:}  During the training process, the training, validation, and testing dataset are split according to each evaluated \BCSD paper. Moreover, similar to existing \BCSD approaches, for a function pair  $(f_{1},f_{2})$, if they are compiled from the same source code, we label their similarity to be 1. Otherwise, we label their similarity to be -1. All the experiments are conducted on a server equipped with  256GB RAM, two Intel CPUs (Xeon R CPU E5-2680, 56 core), and 4 GeForce GTX 2080 GPU cards.

\subsection{Evaluation Datasets}
\label{sub:eval-dataset}
Prior works use different evaluation datasets, making them not directly comparable.  To avoid this problem, we carefully construct two representative  datasets$^1$, which are publicly available.

\textbf{Basic-dataset.} 
%
%
Basic-dataset is used to train the AI-powered \BCSD models and perform similar function detection, consisting of 25 open-source programs, 33 ELFs, and 1,351,104 functions. This dataset involves diverse utilities,  data-processing, and OS-assisting programs,  including OpenSSL, Busybox, Findutils, etc.  We compile these 25 programs with four popular optimization levels (from O0 to O3) under three architectures---ARM, x86, and x64 (supported by over half of the  \BCSD approaches in ~\autoref{tab:BCSD-comparision}).  


\textbf{Application-dataset.}  
%
The application-dataset consists of ten vulnerable functions, querying libraries (OpenSSL and Busybox), and ten IoT firmware images. Specifically, 
the ten vulnerable functions in the widespread third-party library OpenSSL consist of numeric errors, resource management errors, and improper input validation. The ten IoT firmware images involving various utilities, including switch, router, IP camera, and access point, are randomly chosen from four leading IoT vendors (DLink, TPLink, Tuya, and Trendnet). We claim that in this evaluation, we do not plan to perform a large-scale bug or license violation detection to investigate the security state of a large number of IoT firmware images. 
By contrast, we aim to evaluate how the AI-powered \BCSD approaches perform in real-world scenarios. Thus, the ten IoT firmware images which
include 1,935 ELFs and 651,048 functions are enough for this evaluation.

\subsection{Similar Function Detection}
\label{sub:function-search-utilities}
This section performs accuracy and efficiency comparisons of the selected \BCSD approaches. 
According to prior research~\cite{fu2021focus,liu2018alphadiff}, cross-version and cross-compiler analysis of binary snippets compiled with the same architecture and optimization level  is a straightforward \BCSD problem and has been well addressed (since the change of cross-version and cross-compiler binaries is relatively small as discussed in our artifacts). Thus, this paper conducts cross-optimization level and architecture evaluation. 

\begin{table*}
    \scriptsize
    \color{\tablecolor}
	\centering
	\caption{Comparison among AI-powered \BCSD approaches. We rank the experimental results according to precision@5 on mono-architecture under seen programs. We highlight the top-3 experimental results with the best performance. RQ = randomly query result. GQ = the query result after the ground truth change.  NQ = the query result on the new repository. }
	\label{tab:BCSDs-comparision-SS-AUC}\vspace{-6pt}
	\setlength{\tabcolsep}{1.0mm} 
 	{
             \begin{tabular}{lccclllllll}
		\toprule
		\multirow{2}{*}{\textbf{\shortstack{Detector}}}&   \multirow{2}{*}{\textbf{\shortstack{Setting}}}   & \multirow{2}{*}{\textbf{\shortstack{AUC}}}          
            &\multirow{2}{*}{\textbf{\shortstack{ACC}}}& \textbf{\shortstack{Precision@5~\cite{ranking} (\%)}} &\textbf{\shortstack{Recall@5~\cite{ranking}(\%)}}&\textbf{\shortstack{F1-score@5~\cite{ranking} (\%)}}
		&\textbf{\shortstack{Rank-1~\cite{Rank}} (\%)}&  \textbf{\shortstack{MAP@5~\cite{ranking} (\%)}} & \textbf{\shortstack{MRR@5~\cite{MRR} (\%)}}  &\textbf{\shortstack{NDCG@5~\cite{ranking} (\%)}}\\
		  &  &  &   &  RQ/+GQ/NQ   & RQ/+GQ/NQ &  RQ/+GQ/NQ &  RQ/+GQ/NQ &  RQ/+GQ/NQ&  RQ/+GQ/NQ&  RQ/+GQ/NQ\\ 
		\midrule
		
 \rcolor & mono-seen & & & \textbf{44.52}/{\color{black}+}0.79/\textbf{30.71} & \textbf{62.30}/{\color{black}+}0.20/\textbf{38.82} & \textbf{50.56}/{\color{black}+}0.56/\textbf{34.11} & 99.70/{\color{black}+}0.00/\textbf{56.68} & 98.04/{\color{black}+}0.10/\textbf{76.82} & 99.84/{\color{black}+}0.00/\textbf{79.09} & 99.03/{\color{black}+}0.05/\textbf{79.06} \\
 \rcolor & mono-unseen & \multirow{-2}{*}{0.981} & \multirow{-2}{*}{0.930} & 33.25/{\color{black}+}0.19/11.22 & 58.95/{\color{black}+}0.06/16.26 & 40.19/{\color{black}+}0.14/13.14 & \textbf{99.98}/{\color{black}+}0.00/0.67 & 97.49/{\color{black}-}0.02/25.43 & \textbf{99.99}/{\color{black}+}0.00/29.48 & 98.81/{\color{black}-}0.01/30.67 \\
 \rcolor & cross-seen & & & 41.85/{\color{black}+}1.18/30.01 & 25.42/{\color{black}+}0.37/17.16 & 30.29/{\color{black}+}0.66/21.23 & 98.40/{\color{black}+}0.08/52.40 & 97.13/{\color{black}+}0.12/71.84 & 99.29/{\color{black}+}0.03/74.49 & 98.33/{\color{black}+}0.07/74.37 \\ 
\rcolor \multirow{-4}{*}{Gemini-skip~\cite{fu2021focus}} & cross-unseen & \multirow{-2}{*}{0.953} & \multirow{-2}{*}{0.886} & 32.83/{\color{black}+}0.13/13.68 & 23.57/{\color{black}+}0.05/7.95 & 25.48/{\color{black}+}0.07/9.69 & \textbf{99.96}/{\color{black}+}0.00/31.62 & 97.32/{\color{black}-}0.07/37.21 & \textbf{99.98}/{\color{black}+}0.00/41.63 & 98.72/{\color{black}-}0.03/40.95 \\

 & mono-seen & & & \textbf{43.28}/{\color{black}+}0.99/\textbf{31.25} & \textbf{60.96}/{\color{black}+}0.18/\textbf{39.16} & \textbf{49.09}/{\color{black}+}0.64/\textbf{34.55} & 99.90/{\color{black}+}0.00/53.64 & 98.08/{\color{black}+}0.19/\textbf{78.80} & 99.95/{\color{black}+}0.00/\textbf{81.07} & 99.09/{\color{black}+}0.09/\textbf{81.16} \\
 & mono-unseen & \textbf{\multirow{-2}{*}{0.982}} & \multirow{-2}{*}{0.929} & 31.99/{\color{black}+}0.13/7.47 & 57.26/{\color{black}+}0.01/12.00 & 38.75/{\color{black}+}0.09/9.08 & 99.95/{\color{black}+}0.00/0.73 & 97.19/{\color{black}-}0.08/22.18 & 99.98/{\color{black}+}0.00/26.69 & 98.66/{\color{black}-}0.04/25.68 \\
 & cross-seen & & & 39.90/{\color{black}+}1.31/29.89 & 23.87/{\color{black}+}0.40/16.81 & 28.63/{\color{black}+}0.65/20.98 & 98.43/{\color{black}+}0.08/49.94 & 97.53/{\color{black}+}0.13/74.76 & 99.31/{\color{black}+}0.03/77.73 & 98.53/{\color{black}+}0.08/77.42 \\
\multirow{-4}{*}{VulSeeker-skip~\cite{fu2021focus}} & cross-unseen & \multirow{-2}{*}{0.952} & \multirow{-2}{*}{0.877} & 30.50/{\color{black}+}0.15/12.49 & 22.22/{\color{black}+}0.05/6.91 & 23.84/{\color{black}+}0.09/8.68 & \textbf{99.96}/{\color{black}+}0.00/34.00 & 97.65/{\color{black}-}0.08/39.86 & \textbf{99.98}/{\color{black}+}0.00/43.72 & 98.88/{\color{black}-}0.04/43.03 \\

 \rcolor & mono-seen & & & 40.85/{\color{black}+}0.81/28.54 & 58.22/{\color{black}+}0.11/\textbf{35.99} & \textbf{46.53}/{\color{black}+}0.57/\textbf{31.66} & 99.87/{\color{black}+}0.03/49.53 & 98.15/{\color{black}+}0.07/74.47 & 99.92/{\color{black}+}0.02/76.93 & 99.10/{\color{black}+}0.04/76.78 \\
 \rcolor & mono-unseen &\multirow{-2}{*}{0.980}   &\multirow{-2}{*}{0.933}& 32.98/{\color{black}+}0.16/11.23 & 58.61/{\color{black}+}0.03/17.47 & 39.88/{\color{black}+}0.11/13.49 & \textbf{99.98}/{\color{black}+}0.00/0.67 & 97.61/{\color{black}-}0.09/24.92 & 99.98/{\color{black}+}0.00/29.51 & 98.86/{\color{black}-}0.04/29.66 \\
 \rcolor & cross-seen & & & 41.62/{\color{black}+}1.23/28.93 & 24.97/{\color{black}+}0.22/16.50 & 29.84/{\color{black}+}0.52/20.37 & 98.45/{\color{black}+}0.08/51.22 & 96.93/{\color{black}+}0.14/68.37 & 99.28/{\color{black}+}0.04/71.40 & 98.27/{\color{black}+}0.08/71.26 \\
\rcolor \multirow{-4}{*}{Focus-skip~\cite{fu2021focus}} & cross-unseen &\multirow{-2}{*}{0.964}   &\multirow{-2}{*}{0.904} & 33.83/{\color{black}+}0.11/16.56 & 24.26/{\color{black}+}0.02/9.76 & 26.19/{\color{black}+}0.05/11.81 & 99.85/{\color{black}+}0.00/36.88 & 96.92/{\color{black}-}0.04/42.77 & 99.93/{\color{black}+}0.00/47.34 & 98.52/{\color{black}-}0.02/47.04 \\

 & mono-seen & & & 39.11/{\color{black}+}0.74/28.27 & 55.70/{\color{black}+}0.19/35.66 & 44.55/{\color{black}+}0.54/31.36 & 99.00/{\color{black}+}0.00/46.96 & 97.93/{\color{black}+}0.04/76.41 & 99.40/{\color{black}+}0.00/78.54 & 98.83/{\color{black}+}0.02/78.53 \\
 & mono-unseen &\multirow{-2}{*}{0.970}   &\multirow{-2}{*}{0.909}& 33.54/{\color{black}+}0.13/10.50 & 59.36/{\color{black}-}0.01/15.83 & 40.51/{\color{black}+}0.08/12.50 & 99.78/{\color{black}+}0.00/0.67 & 97.25/{\color{black}-}0.03/25.06 & 99.89/{\color{black}+}0.00/30.05 & 98.66/{\color{black}-}0.01/29.81 \\
 & cross-seen & & & 40.55/{\color{black}+}1.65/28.43 & 24.95/{\color{black}+}0.54/16.63 & 29.36/{\color{black}+}0.87/20.26 & 97.60/{\color{black}+}2.20/51.45 & 96.04/{\color{black}+}1.15/70.36 & 98.70/{\color{black}+}1.13/73.09 & 97.63/{\color{black}+}0.92/73.34 \\
\multirow{-4}{*}{Focus~\cite{fu2021focus}} & cross-unseen &\multirow{-2}{*}{0.922} &\multirow{-2}{*}{0.841}& 38.93/{\color{black}+}0.14/{20.04} & 28.94/{\color{black}+}0.03/{13.25} & 30.73/{\color{black}+}0.07/{15.05} & 99.15/{\color{black}+}0.00/45.25 & 95.59/{\color{black}-}0.03/{52.39} & 99.57/{\color{black}+}0.00/{56.39} & 97.79/{\color{black}-}0.02/{56.88} \\

 \rcolor & mono-seen& & & 39.01/{\color{black}+}0.77/27.18 & 55.00/{\color{black}+}0.15/34.30 & 44.36/{\color{black}+}0.52/30.13 & 99.60/{\color{black}+}0.00/45.73 & 97.85/{\color{black}+}0.03/72.22 & 99.77/{\color{black}+}0.00/74.45 & 98.92/{\color{black}+}0.01/74.72 \\
 \rcolor & mono-unseen& \multirow{-2}{*}{0.965}   &\multirow{-2}{*}{0.907} & 31.54/{\color{black}+}0.17/8.61 & 56.64/{\color{black}+}0.04/12.55 & 38.22/{\color{black}+}0.12/10.11 & 99.89/{\color{black}+}0.00/0.50 & 97.37/{\color{black}-}0.09/19.37 & 99.94/{\color{black}+}0.00/24.32 & 98.73/{\color{black}-}0.04/23.36 \\
 \rcolor & cross-seen& & & 44.32/{\color{black}+}1.09/30.83 & 27.10/{\color{black}+}0.31/18.14 & 32.15/{\color{black}+}0.59/22.12 & 98.97/{\color{black}+}0.16/58.19 & 96.68/{\color{black}+}0.15/72.96 & 99.44/{\color{black}+}0.01/75.64 & 98.20/{\color{black}+}0.07/75.84 \\ 
\rcolor \multirow{-4}{*}{Gemini~\cite{xu2017neural}}& cross-unseen   &\multirow{-2}{*}{0.927}   &\multirow{-2}{*}{0.853} & {40.16}/{\color{black}+}0.16/18.08 & {29.96}/{\color{black}+}0.05/12.13 & {31.81}/{\color{black}+}0.09/13.68 & 99.93/{\color{black}+}0.00/40.88 & 95.03/{\color{black}-}0.06/49.19 & 99.97/{\color{black}+}0.00/52.84 & 97.62/{\color{black}-}0.03/53.87 \\

 & mono-seen & & & 36.49/{\color{black}+}1.86/23.17 & 52.02/{\color{black}+}0.09/28.20 & 41.09/{\color{black}+}0.95/24.92 & 99.57/{\color{black}+}0.00/39.54 & 96.97/{\color{black}-}0.01/61.28 & 99.73/{\color{black}+}0.00/64.77 & 98.45/{\color{black}+}0.00/64.85 \\
 & mono-unseen &\multirow{-2}{*}{0.985}   &\textbf{\multirow{-2}{*}{0.944}}& 30.81/{\color{black}+}0.16/9.46 & 55.83/{\color{black}+}0.03/14.19 & 37.44/{\color{black}+}0.11/11.19 & 99.97/{\color{black}+}0.00/0.56 & 97.52/{\color{black}-}0.09/22.48 & \textbf{99.99}/{\color{black}+}0.00/26.54 & 98.82/{\color{black}-}0.04/27.07 \\
 & cross-seen & & & 33.92/{\color{black}+}1.99/23.28 & 18.76/{\color{black}+}0.75/11.38 & 22.82/{\color{black}+}1.12/14.83 & 98.33/{\color{black}+}0.07/38.92 & 97.48/{\color{black}+}0.08/66.03 & 99.49/{\color{black}+}0.05/69.04 & 98.59/{\color{black}+}0.06/69.00 \\
\multirow{-4}{*}{BinaryAI-skipt~\cite{yu2020order}} & cross-unseen &\multirow{-2}{*}{0.969}   &\multirow{-2}{*}{0.915}& 29.67/{\color{black}+}0.12/17.12 & 21.64/{\color{black}+}0.06/9.20 & 23.20/{\color{black}+}0.08/11.74 & \textbf{99.96}/{\color{black}+}0.00/{46.76} & 98.07/{\color{black}-}0.07/52.34 & \textbf{99.98}/{\color{black}+}0.00/56.01 & {99.08}/{\color{black}-}0.03/56.01 \\

 \rcolor & mono-seen& & & 36.01/{\color{black}+}0.61/25.29 & 52.49/{\color{black}+}0.07/32.00 & 41.26/{\color{black}+}0.44/28.11 & \textbf{99.93}/{\color{black}+}0.00/37.78 & 97.67/{\color{black}-}0.06/66.56 & \textbf{99.96}/{\color{black}+}0.00/69.23 & 98.88/{\color{black}-}0.03/69.72 \\
 \rcolor & mono-unseen&\multirow{-2}{*}{0.958}   &\multirow{-2}{*}{0.895} & 30.37/{\color{black}+}0.11/11.05 & 55.18/{\color{black}-}0.01/18.09 & 36.93/{\color{black}+}0.06/13.52 & 99.94/{\color{black}+}0.00/0.73 & 97.09/{\color{black}-}0.03/26.03 & 99.97/{\color{black}+}0.00/30.80 & 98.60/{\color{black}-}0.01/31.16 \\
 \rcolor & cross-seen& & & 34.09/{\color{black}+}1.11/24.40 & 21.32/{\color{black}+}0.30/14.69 & 24.96/{\color{black}+}0.57/17.76 & 97.93/{\color{black}+}1.64/35.36 & 96.94/{\color{black}+}0.62/61.66 & 99.17/{\color{black}+}0.67/64.73 & 98.21/{\color{black}+}0.57/64.94 \\ 
\rcolor \multirow{-4}{*}{VulSeeker~\cite{gao2018vulseeker}}& cross-unseen&\multirow{-2}{*}{0.905}   &\multirow{-2}{*}{0.819}& 32.06/{\color{black}+}0.09/12.92 & 23.60/{\color{black}+}0.02/8.22 & 25.24/{\color{black}+}0.04/9.55 & \textbf{99.96}/{\color{black}+}0.00/29.68 & 96.50/{\color{black}-}0.02/36.84 & \textbf{99.98}/{\color{black}+}0.00/40.78 & 98.32/{\color{black}-}0.01/40.64 \\

 & mono-seen& & & 34.77/{\color{black}+}0.00/22.14 & \textbf{66.51}/{\color{black}+}0.00/30.22 & 42.75/{\color{black}+}0.00/25.34 & 89.93/{\color{black}+}0.00/31.79 & 90.85/{\color{black}+}0.00/58.15 & 93.64/{\color{black}+}0.00/61.36 & 93.52/{\color{black}+}0.00/62.51 \\
 & mono-unseen& \textbf{\multirow{-2}{*}{0.983}}  &\textbf{\multirow{-2}{*}{0.942}}&{33.73}/{\color{black}+}0.00/15.71 & 56.65/{\color{black}+}0.00/21.13 & {40.53}/{\color{black}+}0.00/17.96 & 84.43/{\color{black}+}0.00/0.34 & 87.94/{\color{black}+}0.00/29.40 & 91.07/{\color{black}+}0.00/30.83 & 91.77/{\color{black}+}0.00/39.75 \\
 & cross-seen&  & & \textbf{45.78}/{\color{black}+}0.00/\textbf{32.44} & {35.85}/{\color{black}+}0.00/{20.46} & {35.82}/{\color{black}+}0.00/{23.27} & 91.60/{\color{black}+}0.00/47.91 & 89.93/{\color{black}+}0.00/59.90 & 94.53/{\color{black}+}0.00/64.10 & 93.31/{\color{black}+}0.00/64.88 \\
\multirow{-4}{*}{SAFE~\cite{massarelli2018safe}}& cross-unseen& \multirow{-2}{*}{0.977}  &\multirow{-2}{*}{0.929}&36.36/{\color{black}+}0.00/19.33 & 28.25/{\color{black}+}0.00/12.21 & 28.67/{\color{black}+}0.00/13.98 & 80.11/{\color{black}+}0.00/28.17 & 83.34/{\color{black}+}0.00/42.04 & 87.81/{\color{black}+}0.00/46.68 & 88.27/{\color{black}+}0.00/46.64 \\

 \rcolor& mono-seen & & & 32.50/{\color{black}+}0.55/21.60 & 39.62/{\color{black}+}0.03/31.14 & 33.62/{\color{black}+}0.33/24.24 & 78.93/{\color{black}+}0.22/27.97 & 82.31/{\color{black}+}0.10/48.13 & 85.56/{\color{black}+}0.14/51.98 & 85.90/{\color{black}+}0.11/52.77 \\
 \rcolor& mono-unseen &\multirow{-2}{*}{0.976}   &\multirow{-2}{*}{0.940}& 31.37/{\color{black}+}0.45/{20.83} & 41.46/{\color{black}+}0.03/{31.57} & 33.58/{\color{black}+}0.27/{23.96} & 80.45/{\color{black}+}0.19/{26.41} & 83.67/{\color{black}+}0.09/{48.67} & 86.77/{\color{black}+}0.12/{52.52} & 87.16/{\color{black}+}0.09/{53.47} \\
 \rcolor & cross-seen& & & 29.99/{\color{black}+}0.48/19.05 & 19.11/{\color{black}-}0.01/14.06 & 22.19/{\color{black}+}0.23/15.13 & 61.87/{\color{black}+}0.14/24.62 & 67.80/{\color{black}+}0.06/40.36 & 71.98/{\color{black}+}0.15/44.57 & 72.49/{\color{black}+}0.15/44.88 \\
\rcolor \multirow{-4}{*}{MGMN~\cite{ling2021multilevel}}& cross-unseen & \multirow{-2}{*}{0.961}   &\multirow{-2}{*}{0.899}&28.88/{\color{black}+}0.44/17.94 & 18.73/{\color{black}+}0.01/13.23 & 21.51/{\color{black}+}0.22/14.24 & 61.18/{\color{black}+}0.12/23.11 & 67.60/{\color{black}+}0.05/38.97 & 71.68/{\color{black}+}0.13/43.14 & 72.22/{\color{black}+}0.13/43.32 \\	

 & mono-seen& & & 32.21/{\color{black}+}1.59/18.92 & 46.61/{\color{black}+}0.02/23.36 & 36.60/{\color{black}+}1.04/20.56 & 99.80/{\color{black}+}0.00/28.61 & 96.65/{\color{black}+}0.03/51.25 & 99.88/{\color{black}+}0.00/54.91 & 98.34/{\color{black}+}0.02/55.42 \\
 & mono-unseen& \textbf{\multirow{-2}{*}{0.992}} & \textbf{\multirow{-2}{*}{0.949}} &29.29/{\color{black}+}0.17/8.24 & {53.64}/{\color{black}+}0.06/12.38 & 35.67/{\color{black}+}0.13/9.76 & 99.89/{\color{black}+}0.00/0.44 & 97.36/{\color{black}-}0.09/19.42 & 99.94/{\color{black}+}0.00/22.17 & 98.72/{\color{black}-}0.04/23.02 \\
 & cross-seen& & & 28.24/{\color{black}+}0.91/26.85 & 14.68/{\color{black}+}0.37/14.02 & 18.54/{\color{black}+}0.53/17.66 & 89.35/{\color{black}+}0.86/\textbf{86.02} & 87.35/{\color{black}+}0.56/\textbf{84.87} & 90.80/{\color{black}+}0.72/\textbf{87.96} & 89.37/{\color{black}+}0.59/\textbf{86.84} \\
\multirow{-4}{*}{BinaryAI-bert2~\cite{yu2020order}}& cross-unseen&\textbf{\multirow{-2}{*}{0.983} }  &\multirow{-2}{*}{0.931}& 28.57/{\color{black}+}0.11/11.43 & 20.96/{\color{black}+}0.05/6.10 & 22.45/{\color{black}+}0.07/7.78 & 99.90/{\color{black}+}0.00/28.06 & 97.41/{\color{black}-}0.07/34.20 & 99.95/{\color{black}+}0.00/38.47 & 98.75/{\color{black}-}0.03/37.91 \\

 \rcolor &mono-seen& & & 22.35/{\color{black}+}0.00/8.49 & 48.66/{\color{black}+}0.00/11.44 & 28.39/{\color{black}+}0.00/9.67 & 97.81/{\color{black}+}0.00/4.48 & 97.64/{\color{black}+}0.00/25.13 & 98.78/{\color{black}+}0.00/29.32 & 98.52/{\color{black}+}0.00/28.50 \\
 \rcolor &mono-unseen&\multirow{-2}{*}{0.936}   &\multirow{-2}{*}{0.873}& 20.08/{\color{black}+}0.00/8.57 & 60.46/{\color{black}+}0.00/14.29 & 28.88/{\color{black}+}0.00/10.71 & 98.27/{\color{black}+}0.00/0.40 & 99.09/{\color{black}+}0.00/42.86 & 99.11/{\color{black}+}0.00/47.19 & 99.33/{\color{black}+}0.00/42.86 \\
 \rcolor &cross-seen& & & 23.13/{\color{black}+}0.00/10.34 & 21.33/{\color{black}+}0.00/6.20 & 19.36/{\color{black}+}0.00/7.23 & 98.66/{\color{black}+}0.00/7.45 & 97.88/{\color{black}+}0.00/29.60 & 99.25/{\color{black}+}0.00/33.82 & 98.78/{\color{black}+}0.00/33.26 \\
\rcolor \multirow{-4}{*}{UFE-mean~\cite{Baldoni2018Unsupervised}} &cross-unseen&\multirow{-2}{*}{0.895}   &\multirow{-2}{*}{0.812}& 25.44/{\color{black}+}0.00/10.20 & 20.03/{\color{black}+}0.00/6.01 & 20.19/{\color{black}+}0.00/7.21 & 97.31/{\color{black}+}0.00/8.54 & 96.08/{\color{black}+}0.00/24.13 & 98.63/{\color{black}+}0.00/28.47 & 97.75/{\color{black}+}0.00/28.47 \\

 &mono-seen& && 21.76/{\color{black}+}0.00/11.25 & 49.06/{\color{black}+}0.00/15.41 & 27.90/{\color{black}+}0.00/12.90 & 97.98/{\color{black}+}0.00/4.59 & 98.23/{\color{black}+}0.00/39.35 & 98.88/{\color{black}+}0.00/42.79 & 98.84/{\color{black}+}0.00/42.85 \\
 &mono-unseen&\multirow{-2}{*}{0.929}   &\multirow{-2}{*}{0.884}& 20.32/{\color{black}+}0.00/5.71 & {60.86}/{\color{black}+}0.00/9.52 & 29.18/{\color{black}+}0.00/7.14 & 98.27/{\color{black}+}0.00/0.27 & 98.85/{\color{black}+}0.00/28.57 & 99.11/{\color{black}+}0.00/33.04 & 99.22/{\color{black}+}0.00/28.57 \\
 &cross-seen& & & 21.91/{\color{black}+}0.00/9.64 & 20.61/{\color{black}+}0.00/6.47 & 18.58/{\color{black}+}0.00/7.00 & 98.60/{\color{black}+}0.00/5.07 & 98.34/{\color{black}+}0.00/30.80 & 99.17/{\color{black}+}0.00/34.56 & 98.97/{\color{black}+}0.00/34.10 \\
\multirow{-4}{*}{UFE-attention~\cite{Baldoni2018Unsupervised}}&cross-unseen &\multirow{-2}{*}{0.903}   &\multirow{-2}{*}{0.828}& 24.59/{\color{black}+}0.00/7.88 & 19.46/{\color{black}+}0.00/4.57 & 19.60/{\color{black}+}0.00/5.52 & 97.62/{\color{black}+}0.00/7.78 & 96.47/{\color{black}+}0.00/21.46 & 98.78/{\color{black}+}0.00/26.51 & 97.98/{\color{black}+}0.00/24.69 \\

 \rcolor &mono-seen & & & 20.86/{\color{black}+}0.00/9.39 & 47.64/{\color{black}+}0.00/12.47 & 26.82/{\color{black}+}0.00/10.65 & 98.89/{\color{black}+}0.00/1.59 & \textbf{98.85}/{\color{black}+}0.00/27.53 & 99.39/{\color{black}+}0.00/30.96 & \textbf{99.28}/{\color{black}+}0.00/32.13 \\
 \rcolor  &mono-unseen &\multirow{-2}{*}{0.838}   &\multirow{-2}{*}{0.702}&20.05/{\color{black}+}0.00/8.57 & 60.42/{\color{black}+}0.00/14.29 & 28.85/{\color{black}+}0.00/10.71 & 98.27/{\color{black}+}0.00/0.13 & \textbf{99.11}/{\color{black}+}0.00/26.19 & 99.11/{\color{black}+}0.00/30.18 & \textbf{99.34}/{\color{black}+}0.00/30.44 \\
 \rcolor &cross-seen & & & 21.07/{\color{black}+}0.00/7.62 & 20.49/{\color{black}+}0.00/5.00 & 18.18/{\color{black}+}0.00/5.60 & {99.12}/{\color{black}+}0.00/2.74 & \textbf{99.02}/{\color{black}+}0.00/23.27 & {99.52}/{\color{black}+}0.00/27.38 & \textbf{99.40}/{\color{black}+}0.00/26.14 \\
\rcolor \multirow{-4}{*}{UFE-rnn~\cite{Baldoni2018Unsupervised}} &cross-unseen &\multirow{-2}{*}{0.691}   &\multirow{-2}{*}{0.548}& 21.07/{\color{black}+}0.00/5.45 & 17.91/{\color{black}+}0.00/3.47 & 17.36/{\color{black}+}0.00/3.97 & 98.35/{\color{black}+}0.00/1.41 & {98.32}/{\color{black}+}0.00/12.76 & 99.15/{\color{black}+}0.00/17.12 & 98.96/{\color{black}+}0.00/16.07 \\

 &mono-seen&  &  & 2.00/+0.00/2.00 &36.11/+0.00/33.33&3.78/++0.00/3.72&8.33/++0.00/2.00&12.38/++0.00/23.51&12.38/++0.00/23.51&27.9/+0.00/37.26\\
\multirow{-2}{*}{Asm2Vec~\cite{ding2019asm2vec}} &mono-unseen& \multirow{-2}{*}{0.703} & \multirow{-2}{*}{0.300}&2.00/+0.00/2.21&4.59/+0.00/3.62&3.81/+0.00/4.13&1.46/+0.00/1.45&8.08/+0.00/6.54&8.12/+0.00/7.19&25.05/+0.00/24.67\\

\rcolor  &mono-seen & 0.811 & 0.496&-   &- &- &- &- &- &- \\
\rcolor \multirow{-2}{*}{ Asteria~\cite{yang2021asteria}} &cross-seen & 0.874 & 0.496&-   &- &- &- &- &- &- \\
	\bottomrule
	\end{tabular}}
\end{table*}

\subsubsection{Accuracy Evaluation}
\label{subsub:EvaluationacrossArchitectures}
We first perform cross-optimization level and architecture comparison. Then, we investigate two factors that may affect the performance of a \BCSD tool.  
Specifically, according to our empirical analysis, 
 prior researchers usually label a repository function in the top-K search list as a true positive if the repository function has the same name as the query function. Otherwise, the repository function is a false positive. However, sometimes the repository function is just renamed and is still compiled from the same source code as the query function. In this condition, the repository function should be considered as a true positive. Thus, it is necessary to investigate how function renames influence similar function detection.  
 Moreover,  \BCSD performance is also influenced by how we construct the searching repository. Specifically, suppose a repository contains functions very similar to the query function, i.e., the binary change between the query and repository functions is small (for instance, when the query and repository functions are compiled with -O2 and -O3 optimization levels, respectively). In that case,  a \BCSD tool can easily recognize the corresponding similar functions and achieve good performance. Thus, we also need to explore how the constructed searching repository affects \BCSD performance.

\textbf{Cross-optimization level and architecture evaluation.} 
As shown in~\autoref{tab:BCSDs-comparision-SS-AUC}, we compare existing AI-powered \BCSD approaches by using 2 common machine learning metrics and 7 widely used ranking metrics in recommendation systems on both the seen and unseen datasets (the seen dataset is used to train embedding models). 
Each metric value in the RQ columns is the average of 3,000 times random function queries. For each query, the repository consists of 100,000  randomly selected functions. The function comparison process of Asteria~\cite{yang2021asteria} is extremely slow (several hours per function search). We fail to perform 3000 times of queries for this approach. Thus, we do not report its ranking metrics.

By analyzing these experimental results, we obtain the following findings. 
(1) No one \BCSD approach always performs the best on all metrics. For instance, BinaryAI-bert2 achieves the best AUC, while Gemini-skip performs better in ranking metrics. This experimental result indicates that different \BCSD approaches may be applicable in different scenarios.
(2) The GNN-based \BCSD approaches achieve top-level ranking metrics in the current literature,  which demonstrates GNN is a promising embedding neural network in \BCSD. 
(3) Compared to mono-ISA, recall values decrease significantly in cross-ISA. For instance, the recall@5 value of Gemini-skip decreases from 62.3\% to 25.42\% when the evaluation setting changes from mono-seen (the evaluation on mono-ISA and seen dataset) to cross-seen (the evaluation on cross-ISA and seen dataset). This experimental result demonstrates that \BCSD models tend to identify similar functions compiled under the same ISA. Thus, {cross-architecture \BCSD still needs further improvement.}
(4) Most \BCSD approaches perform well on ranking metrics, including Rank-1, MAP, MRR, and NDCG. Furthermore, for most \BCSD methods, these metric values are similar, which means sometimes, using these metrics can hardly differentiate excellent \BCSD approaches. 
(5) A higher AUC value does not mean a better capability of similar function detection.   For instance, though BinaryAI-bert2 achieves the best AUC (99.2\%) and ACC (94.9\%), its precision (32.21\%) is 12.3\% less than Gemini-skip. Besides, the AUC values of many \BCSD approaches are similar. Thus, only using the AUC value can hardly present the capability of a \BCSD tool.
(6) The \BCSD systems that use instruction embeddings (obtained by skip-gram model~\cite{Mikolov2013Distributed}) as the basic block features perform better than the ones using manually extracted features. 
For instance, both the AUC and precision of Focus-skip and VulSeeker-skip are better than Focus and VulSeeker, respectively. This experimental result demonstrates that neural networks are effective in extracting representative binary features in code embedding. 
Next, we explore key factors affecting the performance of each \BCSD approach.

\textbf{The influence of function rename.}  
Instead of function name comparison, we identify all the real TPs in the function search list by comparing the source code of the search results with query functions. After we rectify the FPs (compiled by the same source code as the query function while having different function names) to TPs, we show the metric value changes in the GQ columns in~\autoref{tab:BCSDs-comparision-SS-AUC}. We observe that the prior ground truth labeling method decreases the precision and recall
of \BCSD approaches. For instance, after rectification, the precision of Gemini-skip increased by 1.18\%. Fortunately, the influence of function renames is limited. 
 \begin{figure}[]
        \centering
            \subfloat[BinaryAI-bert2]{
                \label{ratio-bert-ordermatters} 
                \includegraphics[width=0.2\textwidth]{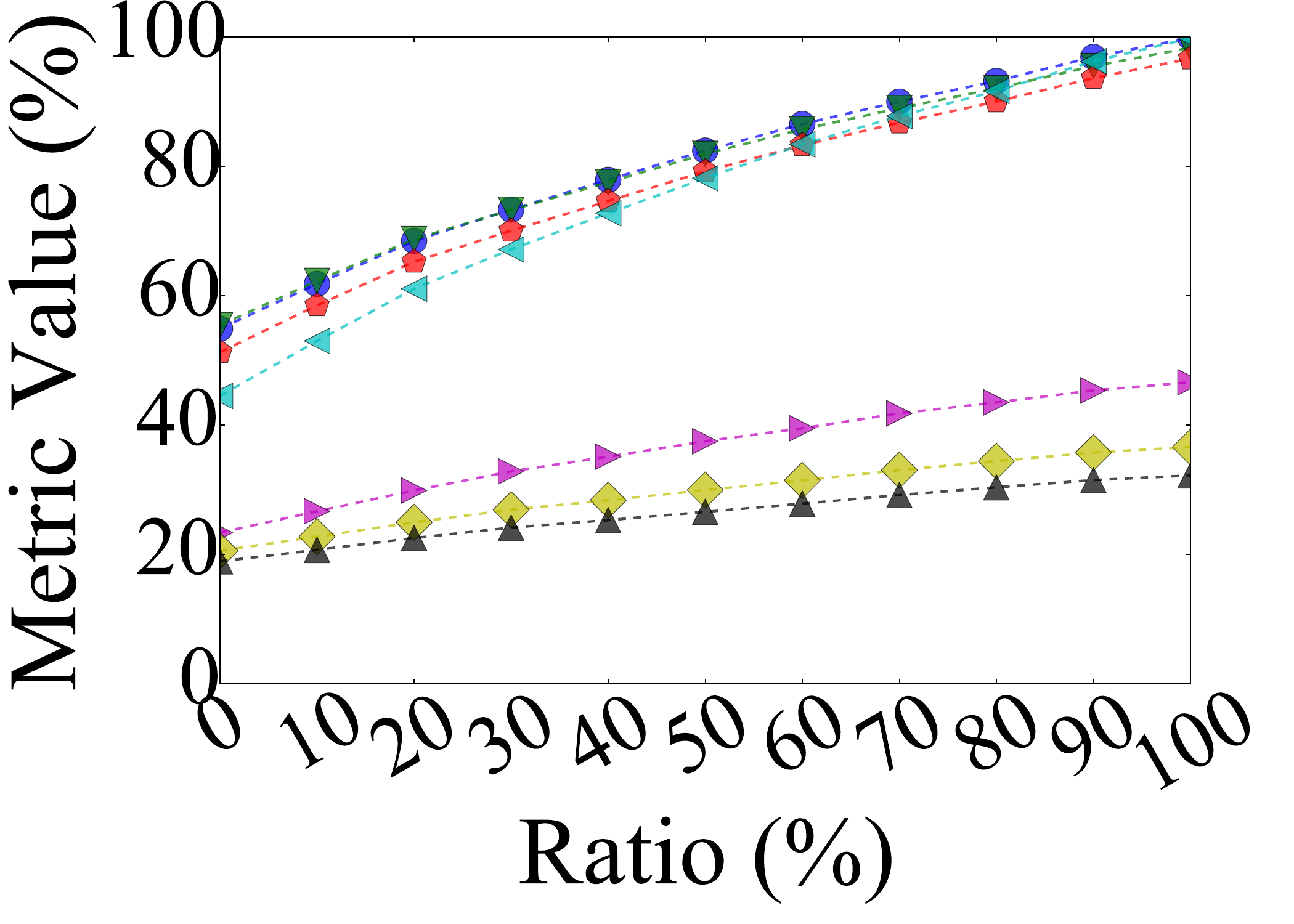}}
            \subfloat[Gemini-skip]{
                \label{ratio-i2v-gemini} 
                \includegraphics[width=0.25\textwidth]{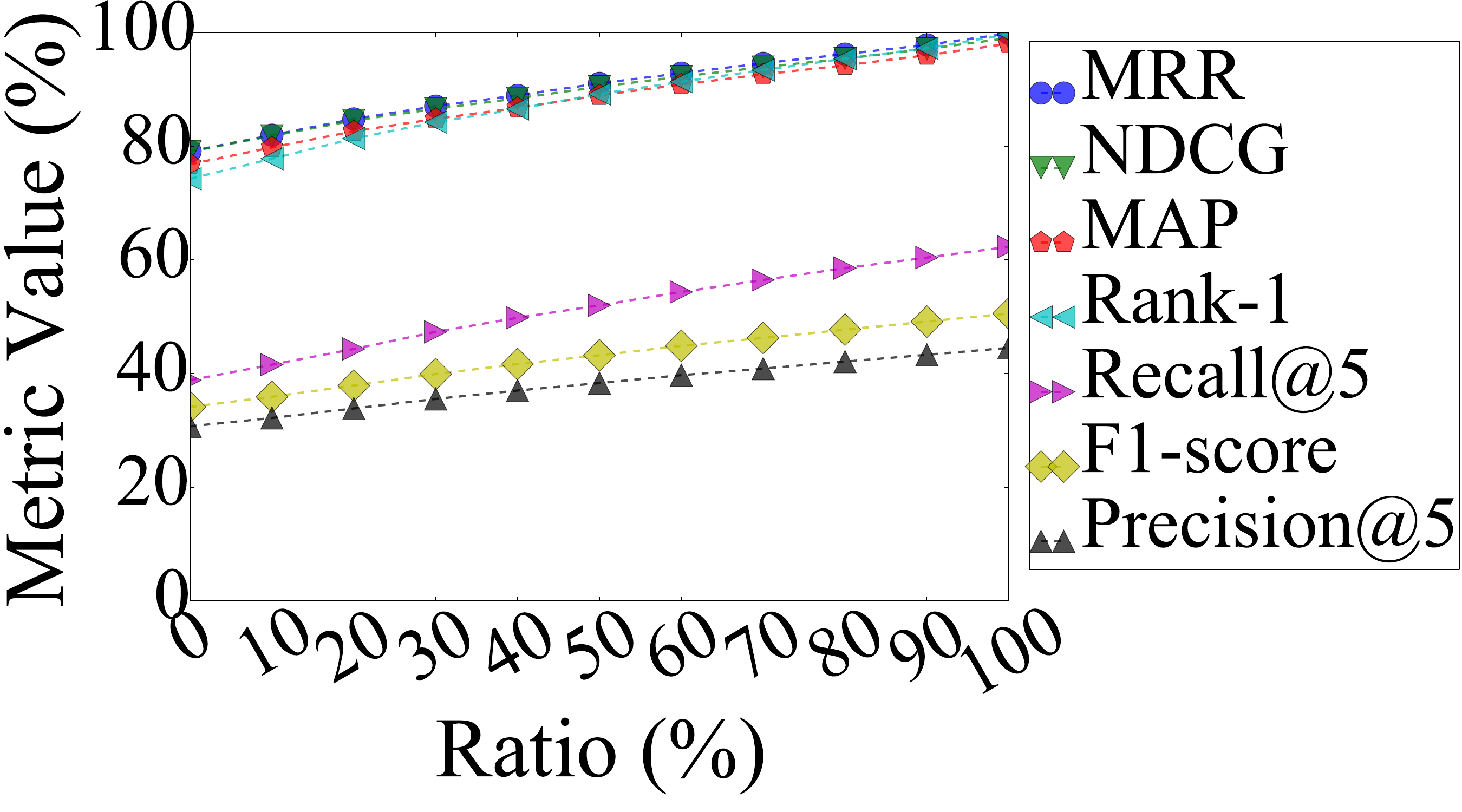}}
            
        \caption{We change the way of building the searching repository by including different ratios of query functions in the constructed repository. (a) and (b) show how metric values change under different ratios.}
        \label{fig:metrics-change-with-ratio}
\end{figure}

 \textbf{The influence of repository construction.}
According to our empirical analysis, the same functions as the query function in the repository are usually identified at the top-1 ranking, which significantly influences the value of most position-sensitive metrics in recommendation systems such as MRR, MAP, and NDCG. To investigate how repository construction affects the performance of \BCSD approaches, after randomly generating repository functions, we intentionally remove the functions that are the same as the query functions to construct a new repository. Then, we perform similar function detection by using the newly built repository. The metric values obtained by querying the new repository are shown in the NQ columns in~\autoref{tab:BCSDs-comparision-SS-AUC}. 
After removing the same functions as query functions from the constructed repository, ranking metrics such as Rank-1, MAP, and MRR decrease significantly. For instance,  the Rank-1 value of BinaryAI-bert2 drops from 99.80\% to only 28.61\%  on mono-architecture evaluation. Moreover, we further investigate how repository construction influences evaluation metrics by involving different ratios of query functions in the
constructed repository. The experimental result shown  in~\autoref{fig:metrics-change-with-ratio} indicates that how we build the searching repository significantly influences the ranking metrics of \BCSD---the more the repository contains the query functions, the better the evaluation metrics we obtain. 
 
\textbf{Summary of findings.} We summarize how most metrics change under different conditions in~\autoref{tab:MainFindings} to clearly present the findings in accuracy evaluation. For example, when the evaluated dataset changes from the seen dataset to the unseen dataset, the AUC and precision decrease a little, the recall decreases significantly, and other ranking metrics do not change much. Particularly, the accuracy of similar function detection is significantly influenced by the value of K in ranking metrics~\cite{valcarce2020assessing}. Thus, we also measure the impact of K. More comparisons can be found in our artifacts.

\subsubsection{Efficiency Comparision}
\label{subsub:Efficiencycomparision}
 This section presents the efficiency comparison of  state-of-the-art AI-powered \BCSD tools. Considering that (1) the preprocessing and code representation phases of many  \BCSD tools are similar, and (2) the efficiency of these two phases has been well described in prior research. For instance, Focus~\cite{fu2021focus} demonstrates the extraction time of CFG is within 0.05s per function. Thus, this section evaluates the training and function embedding time of \BCSD approaches, which are influenced by the structures of neural networks.
 We perform this evaluation under single and cross-platforms using the training data. 
 As shown in ~\autoref{tab:BCSDs-comparision-efficiency}, the training and function embedding time of each \BCSD approach vary a lot because different \BCSD approaches are composed of different neural network structures. For instance, it takes 36 minutes for MGMN to train a mono-architecture model. By contrast, Asteria requires 23 hours and 33 minutes to finish the training process. Moreover, the difference in the function embedding time is even larger. It takes 336 hours for Focus-skip to generate function embeddings on cross-platforms. By contrast, Gemini only needs 1 hour and 31 minutes to finish the embedding generation process. Considering that model training is a one-time effort work, researchers should focus on the embedding time, which directly influences the practicability of a \BCSD tool in real-world scenarios.

\subsection{Downstream Application }
\label{sub:downstreamapplication}


\begin{table}
    \scriptsize
    \color{\tablecolor}
	\centering
 \caption{Summary of findings in this section. $\downarrow$ = decrease. $\Downarrow$ = decrease significantly. $\uparrow$ = increase. $\Uparrow$ = increase significantly. $\rightsquigarrow$ = does not change much. - = unrelated to the condition. }
	
	\label{tab:MainFindings}\vspace{-6pt}
	\setlength{\tabcolsep}{1.0mm} 
 	{
	\begin{tabular}{clccccccc}
		\toprule
		\textbf{ID} &\textbf{\shortstack{Condition}} &\textbf{AUC}& \textbf{\shortstack{Precision}} & \textbf{\shortstack{Recall}}& \textbf{Rank-1}&\textbf{MAP} & \textbf{MRR}& \textbf{NDCG}\\
		\midrule
           1&Mono -> Cross & $\downarrow$ & $\downarrow$ &$\Downarrow$ & $\rightsquigarrow$ & $\rightsquigarrow$ & $\rightsquigarrow$ & $\rightsquigarrow$\\ 
           2&Seen -> Unseen & - &$\Downarrow$& $\downarrow$ & $\rightsquigarrow$ & $\rightsquigarrow$ & $\rightsquigarrow$ & $\rightsquigarrow$\\ 
           3& MFE -> NLP  & $\uparrow$  & $\uparrow$& $\uparrow$& $\rightsquigarrow$ & $\rightsquigarrow$ & $\rightsquigarrow$ & $\rightsquigarrow$\\ 
           4& K\_Value $\hookuparrow$ & -&$\Downarrow$ & $\uparrow$ &$\rightsquigarrow$ &$\downarrow$&$\downarrow$ &$\downarrow$\\ 
           5& Query\_Function\_Ratio $\hookuparrow$ &- &$\uparrow$&$\uparrow$ &$\uparrow$&$\uparrow$&$\uparrow$&$\uparrow$ \\  
		\bottomrule
	\end{tabular}}
\end{table}

\subsubsection{Vulnerability Search}
\label{subsub:VulnerabilitySearch}
This section evaluates whether the selected \BCSD approaches can identify the ten vulnerabilities existing in IoT firmware images.  
To facilitate the vulnerability detection process, we compile the ten vulnerabilities under ARM, the same platform as the chosen IoT firmware images. Moreover, most IoT firmware images are compiled with -O2/-O3 optimization levels. To increase the detection accuracy, we compile the ten vulnerabilities with -O2 and -O3 optimization levels. Finally, we obtain 20 vulnerable functions. Next, conduct the vulnerability search and obtain a list of search results that \BCSD systems report being similar vulnerabilities in IoT firmware images. To check whether the top-K (we choose K to be 10 to identify as many vulnerable functions as possible) search results are real vulnerabilities, we first examine whether the executable file name of the search result is identical to the one of the query function. If so, we further check the stripped binary functions to determine whether it is a bug by manually comparing the pseudocode snippets (an ISA agnostic language similar to source code) of the search result and query function obtained by IDA pro~\cite{idaPro70}. For instance, we manually check and compare similar function bodies such as local variables, program loops, call functions, and constant strings.

\begin{table}
     
    \scriptsize
	\centering
	\caption{Efficiency comparison of AI-powered \BCSD approaches. Asm2Vec does not support cross platform \BCSD. MGMN and Asteria are end-to-end methods. Thus, we cannot calculate their function embedding time. 
        }
	\label{tab:BCSDs-comparision-efficiency}\vspace{-6pt}
	\setlength{\tabcolsep}{1.6mm} 
 	{
	\begin{tabular}{lllll}
		\toprule
		\multirow{2}{*}{\textbf{Detector}}  & \multicolumn{2}{c}{\textbf{\shortstack{Training Time}}} & \multicolumn{2}{c}{\textbf{\shortstack{Embedding Time}}} \\ 
		\cmidrule(lr){2-3}
		\cmidrule(lr){4-5}
		& \textbf{\shortstack{x86}} &\textbf{\shortstack{cross}}& \textbf{\shortstack{x86}} &\textbf{\shortstack{cross}}\\
		\midrule
        \rcolor MGMN~\cite{ling2021multilevel}  & 36m   &1h42m & - & -\\
        SAFE~\cite{massarelli2018safe}  & 1h12m   &3h1m & 35m & 4h11m \\ 
        \rcolor Asm2Vec~\cite{ding2019asm2vec}  & 1h27m   &- & 3h27m & -\\
		Gemini~\cite{xu2017neural}  & 2h05m   &2h06m & 32m  & 1h31m \\ 
		\rcolor Gemini-skip~\cite{fu2021focus}  & 2h56m   &2h49m & 43m & 1h29m\\ 
		VulSeeker~\cite{gao2018vulseeker}  & 5h34m   &5h44m & 2h12m & 6h7m\\ 
		\rcolor BinaryAI-bert2~\cite{yu2020order}  & 5h44m   &5h29m & 4h56m & 10h1m\\ 
		BinaryAI-skipt~\cite{yu2020order}  & 6h14m   &9h38m& 5h9m & 8h21m \\ 
		\rcolor VulSeeker-skip~\cite{fu2021focus}  & 6h27m   &5h33m & 3h25m & 5h59m \\ 
		UFE-mean~\cite{Baldoni2018Unsupervised}  & 9h46m   &25h30m & 3h6m & 16h51m\\ 
		\rcolor UFE-attention~\cite{Baldoni2018Unsupervised}  & 10h7m   &24h50m & 3h7n & 16h15m\\
		Focus-skip~\cite{fu2021focus}  & 11h32m   & 18h56m  & 23h48m & 336h38m\\
		\rcolor UFE-rnn~\cite{Baldoni2018Unsupervised}  & 12h40m   & 16h43m & 3h14m & 16h43m\\ 
		Focus~\cite{fu2021focus}  & 17h44m   &17h51m & 23h49m & 263h42m\\
		\rcolor Asteria~\cite{yang2021asteria} &22h33m&94h29m&-&-\\
		\bottomrule
	\end{tabular}}\vspace{-10pt}
\end{table}

\PP{Findings}
From the vulnerability search result shown in~\autoref{tab:IoT-Bug}, we find that: (1) Gemini-skip achieves the best performance (detecting 24 vulnerabilities in the ten IoT firmware images). This experimental result is consistent with the one shown in similar function detection---GNN-based \BCSD approaches are more applicable in the problem of \BCSD;  
(2) The false positives need further improvement for the evaluated \BCSD approaches. For instance, for all the search results reported by UFE-attention, only four of them are vulnerable functions. In this condition, it requires great manual effort to identify the real bugs among the reported search results. 
(3) It is infeasible to use a certain threshold to determine whether a search result is a real vulnerability or not. For instance, when querying the CVE-2014-0195,  Gemini-skip reports seven bugs with maximum and minimum similarity scores of 0.93 and 0.8, respectively. In this condition, the threshold should be lower than 0.8 to ensure all the vulnerabilities are reported.
However, when querying CVE-2014-3513, the similarity scores of all the top 10 results are higher than 0.9. In this condition, Gemini-skip is supposed to report all ten search results to be vulnerable. Unfortunately, none of them are true vulnerabilities, meaning that using a threshold to determine semantically similar functions is impractical.
In summary, the current \BCSD approaches still need more improvement to perform vulnerability detection. 

\begin{table*}
    
    \scriptsize
    \color{\tablecolor}
	\centering
	\caption{Vulnerability detection.}
	 
	\label{tab:IoT-Bug}\vspace{-6pt}
	\setlength{\tabcolsep}{0.6mm} 
 	{
	\begin{tabular}{cccccccccccccccc}
		\toprule
		\multirow{2}{*}{\textbf{\shortstack{CVE}} }  & \multicolumn{3}{c}{\textbf{\shortstack{Gemini-skip~\cite{fu2021focus}}}} &\multicolumn{3}{c}{\textbf{\shortstack{BinaryAI-bert2~\cite{yu2020order}}}} & \multicolumn{3}{c}{\textbf{\shortstack{VulSeeker-skip~\cite{gao2018vulseeker}}}} & \multicolumn{3}{c}{\textbf{\shortstack{SAFE~\cite{massarelli2018safe}}}} & \multicolumn{3}{c}{\textbf{\shortstack{UFE-attention~\cite{Baldoni2018Unsupervised}}}} \\
		\cmidrule(lr){2-4} \cmidrule(lr){5-7} \cmidrule(lr){8-10} \cmidrule(lr){11-13} \cmidrule(lr){14-16}
		 & \textbf{\shortstack{Max\\Similarity}}   & \textbf{\shortstack{Min\\Similarity}} &  \textbf{\shortstack{Bugs}}  & \textbf{\shortstack{Max\\Similarity}}  & \textbf{\shortstack{Min\\Similarity}} &  \textbf{\shortstack{Bugs}}  & \textbf{\shortstack{Max\\Similarity}}  & \textbf{\shortstack{Min\\Similarity}} &  \textbf{\shortstack{Bugs}} & \textbf{\shortstack{Max\\Similarity}}  & \textbf{\shortstack{Min\\Similarity}} &  \textbf{\shortstack{Bugs}} & \textbf{\shortstack{Max\\Similarity}}  & \textbf{\shortstack{Min\\Similarity}} &  \textbf{\shortstack{Bugs}} \\
		\midrule
        \rcolor CVE-2014-3571 & 0.97 & 0.9 & 0 & 0.97 & 0.85 & 4 & 0.99 & 0.96 & 0 & 0.31 & 0.21 & 4 & 0.97 & 0.88 & 2 \\
        CVE-2015-0289 & 0.95 & 0.83 & 0 & 0.99 & 0.77 & 0 & 0.98 & 0.92 & 0 & 0.32 & 0.2 & 0 & 0.95 & 0.82 & 1 \\
        \rcolor CVE-2016-2181 & 0.96 & 0.88 & 0 & 0.98 & 0.88 & 0 & 0.98 & 0.9 & 0 & 0.37 & 0.31 & 0 & 0.98 & 0.91 & 0 \\
        CVE-2015-1791 & 0.95 & 0.84 & 0 & 0.99 & 0.82 & 0 & 0.98 & 0.9 & 0 & 0.48 & 0.33 & 0 & 0.93 & 0.79 & 0 \\
        \rcolor CVE-2015-1792 & 0.95 & 0.87 & 1 & 0.99 & 0.91 & 0 & 1.0 & 0.96 & 2 & 0.41 & 0.32 & 0 & 0.96 & 0.82 & 0 \\
        CVE-2014-3513 & 0.98 & 0.9 & 0 & 0.98 & 0.88 & 0 & 0.98 & 0.91 & 0 & 0.35 & 0.27 & 0 & 0.98 & 0.88 & 0 \\
        \rcolor CVE-2014-0160 & 0.92 & 0.77 & 0 & 0.97 & 0.82 & 0 & 0.98 & 0.91 & 0 & 0.37 & 0.27 & 0 & 0.96 & 0.84 & 0 \\
        CVE-2016-0797 & 0.97 & 0.88 & 8 & 0.94 & 0.7 & 6 & 0.98 & 0.93 & 6 & 0.3 & 0.24 & 0 & 0.97 & 0.88 & 0 \\
        \rcolor CVE-2016-2176 & 0.98 & 0.9 & 8 & 0.95 & 0.74 & 0 & 0.98 & 0.91 & 0 & 0.41 & 0.34 & 0 & 0.93 & 0.78 & 1 \\
        CVE-2014-0195 & 0.93 & 0.8 & 7 & 0.93 & 0.76 & 4 & 0.97 & 0.9 & 2 & 0.41 & 0.27 & 0 & 0.97 & 0.81 & 0 \\
        \bottomrule
	\end{tabular}}
\end{table*}

\begin{table*}
    \scriptsize
    \color{\tablecolor}
	\centering
	\caption{License violation detection (the rank of the target program that is similar to the query program).  ``-1" = the query ELF does not exist in the       
                 firmware; ``bb" = Busybox;  ``lc" = libcrypto; ``ls" = libssl; ``op" = openssl. }
	\label{tab:Code-plag}\vspace{-6pt}
	\setlength{\tabcolsep}{0.9mm} 
 	{
	\begin{tabular}{lccccccccccccccccccccc}
		\toprule
		\multirow{2}{*}{\textbf{\shortstack{Firmware}}} & \multirow{2}{*}{\textbf{\shortstack{\#ELF}}} & \multicolumn{4}{c}{\textbf{\shortstack{Gemini-skip~\cite{fu2021focus}}}} & \multicolumn{4}{c}{\textbf{\shortstack{BinaryAI-bert2~\cite{yu2020order}}}} & \multicolumn{4}{c}{\textbf{\shortstack{SAFE~\cite{massarelli2018safe}}}} & \multicolumn{4}{c}{\textbf{\shortstack{UFE-attention~\cite{Baldoni2018Unsupervised}}}} & \multicolumn{4}{c}{\textbf{\shortstack{VulSeeker-skip~\cite{gao2018vulseeker}}}}  \\
        \cmidrule(lr){3-6} \cmidrule(lr){7-10} \cmidrule(lr){11-14} \cmidrule(lr){15-18} \cmidrule(lr){19-22}
        &&bb&lc&ls&sl&bb&lc&ls&op&bb&lc&ls&op&bb&lc&ls&op&bb&lc&ls&op \\
        \midrule
         
        CAP1200v1_1.0.0_20170801-rel61314_up & 176 & 5 & 1 & 5 & -1 & 1 & 1 & 2 & -1 & 11 & 3 & 14 & -1 & 3 & 147 & 67 & -1 & 6 & 1 & 8 & -1\\
        \rcolor COM_T01F001_LM.1.6.18P12_sign2_TPL.TL-SC4171G & 97 & 2 & 1 & 5 & -1 & 1 & 1 & 2 & -1 & 3 & 1 & 4 & -1 & 106 & 10 & 36 & -1 & 2 & 1 & 4 & -1\\
        COM_T01F001_LM.1.6.18P7_TPL.TL-SC4171G & 96 & 2 & 1 & 5 & -1 & 3 & 1 & 2 & -1 & 3 & 1 & 4 & -1 & 104 & 7 & 82 & -1 & 2 & 1 & 4 & -1\\
        \rcolor COVR-2600R_FW101b05_0911_txbfdisable0911190427 & 281 & 5 & 1 & 4 & 1 & 1 & 1 & 2 & 3 & 17 & 4 & 15 & 166 & 2 & 233 & 230 & 16 & 7 & 1 & 7 & 11\\
        COVR-2600R_FW101b05_beta01_hcr2 & 284 & 5 & 1 & 4 & 1 & 1 & 1 & 2 & 3 & 16 & 4 & 14 & 167 & 267 & 282 & 284 & 225 & 7 & 1 & 7 & 11\\
        \rcolor COVR-3902_ROUTER_v101b05 & 284 & 5 & 1 & 4 & 1 & 1 & 1 & 2 & 3 & 16 & 4 & 14 & 167 & 202 & 256 & 180 & 77 & 7 & 1 & 7 & 11\\
        DAP2610-firmware-v101-rc017 & 128 & 6 & 1 & 4 & 3 & 1 & 2 & 3 & 4 & 16 & 3 & 10 & 76 & 8 & 126 & 86 & 54 & 8 & 2 & 5 & 11\\
        \rcolor DLINK_DNR-322.2.10b022.10.0612.2014 & 236 & 17 & 4 & 20 & 33 & 8 & 1 & 2 & 23 & 26 & 17 & 40 & 96 & 9 & 171 & 47 & 35 & 7 & 4 & 19 & 31\\
        DLINK_DNR-322L.1.40b011.16.1219.2012 & 225 & 10 & 7 & 13 & 24 & 7 & 1 & 2 & 22 & 19 & 10 & 30 & 84 & 174 & 47 & 160 & 28 & 7 & 4 & 17 & 28\\
        \rcolor Dap2610-firmware-v101-beta28-rc0480306165616 & 128 & 6 & 1 & 4 & 3 & 1 & 2 & 3 & 4 & 16 & 2 & 10 & 75 & 7 & 93 & 29 & 15 & 8 & 2 & 5 & 10\\
		\bottomrule
	\end{tabular}}
\end{table*}

\subsubsection{License Violation Detection}
\label{subsub:CodePlagiarismDetection}
This section checks whether the evaluated \BCSD approaches can recognize the GPL-licensed third-party libraries such as OpenSSL in the closed-source IoT firmware images.
We choose 4 ELFs in OpenSSL-1.0.1f and Busybox-1.27.0 as the query libraries since they are widely used in IoT firmware images. Considering a library consists of many functions, and the similarity of two libraries is closely related to their function similarity,  we average the similarity score ($s_{f}$) of all the query functions in a query library $Q$ as the similarity score $S_{QT}$ between the query library  $Q$ and the target library $T$. 
Specifically, for each query function $f$ in the query library, we search for it in each target library $T$ and record its top-1 similarity score $s_{f}$. Then, we average all the $s_{f}$ to obtain $S_{QT}$. 
In this way, we compare the query library with all the target libraries in the IoT firmware images, sort all target libraries according to $S_{QT}$, and record the rank of the target library that is similar to the query library.

\PP{Findings}
From the experimental result shown in ~\autoref{tab:Code-plag}, we find that (1) AI-powered \BCSD approaches achieve relatively good performance in license violation detection. Specifically, BinaryAI-bert2, Gemini-skip, VulSeeker, SAFE, and UFE-attention recognize 15, 11, 6, 2, and 0 query ELFs at the top-1 position. 
(2) Among all the evaluated \BCSD systems, BinaryAI-bert2 can identify the most GPL libraries in IoT firmware images. 
We conclude the root reason is that in license violation detection, the search repository for each function is relatively small (compared to the task of vulnerability search, the repository functions in this task consist of the functions in the target library and are usually no more than several thousand). In this condition, the adjacency matrix of CFG is a good feature to characterize a binary function since there exist fewer different functions that happen to have similar adjacency matrices as the query function in a relatively small repository. 
(3) Compared to vulnerability detection, the evaluated five \BCSD approaches can detect program similarity with acceptable accuracy. For instance, BinaryAI-bert2 can report most target libraries (the one similar to the query library) within top-3 search results. The reason is that vulnerability detection requires accurate \BCSD detection for each function pair, which is still challenging for existing \BCSD approaches. By contrast, license violation detection is based on the statistical \BCSD detection results of all the functions in the query program. 
Thus, existing \BCSD approaches can achieve good performance in license violation detection.

\section{Understanding of Embedding Networks and Evaluation Metrics}

As shown in~\autoref{phaseofBCSD}, \BCSD approaches first perform code disassembly and feature extraction and then use embedding neural networks to obtain low-dimensional vector representation. This evaluation does not evaluate the impact of disassembly tools since Zhou. et al.~\cite{jiang2022comprehensive} have performed a comprehensive study on disassembly tools, demonstrating that IDA Pro~\cite{idaPro70} has the highest precision on many disassembly tasks. Moreover, Marcelli et al.~\cite{marcellimachine} have compared feature extraction methods and provided insightful findings. Thus, this section  next investigates embedding neural networks and the widely adopted evaluation metrics in \BCSD approaches. 

\subsection{ Embedding Neural Networks}
\label{sbu:embeddingneuralnetworks}
We perform a structure analysis together with the trace of neuron output to understand neural networks. 
It is well known that model understanding is a rather difficult task. Thus, we do not investigate all the embedding networks listed in~\autoref{tab:BCSD-comparision}. By contrast, we explore two representative embedding neural networks (GNN and CNN) that have been widely used in \BCSD. We hope our investigation can inspire future work in this field.

\subsubsection{GNN}
\label{analysis-of-GNN}
Many \BCSD approaches~\cite{xu2017neural,fu2021focus,gao2018vulseeker,ling2021multilevel,Baldoni2018Unsupervised} propose to use GNN to embed binary functions. We observe that all GNNs, including the Structure2Vec~\cite{dai2016discriminative}, DNN~\cite{gao2018vulseeker}, GAT~\cite{velickovic2017graph} used in existing  \BCSD systems, follow a neighborhood aggregation strategy---iteratively updating the representation of a vertex by aggregating the representations of its neighbors. In this way, after $T$ iterations of aggregation, the representation of a vertex can contain the structural information within its H-hop neighborhood. 
Specifically, for a vertex $i$, GNN first aggregates the representations of its neighbors in iteration $t$ to be $\vec{h}_i^{t+1}$ (as shown in~\autoref{AGGREGATE}). Then, GNN represents the vertex embedding in iteration $t$ as $\vec{\mu}_i^{t+1}$ by a $Union$ function (as shown in~\autoref{Union}). Finally, in~\autoref{READOUT}, by using a $READOUT$ function, GNN aggregates all the vertex embeddings to form the entire graph's representation $\vec{h}_G$. 
Such a neighborhood aggregation strategy means that \textit{a maximally powerful GNN would never map two different neighborhoods, i.e., multi-sets of feature vectors, to the same representation.}   Therefore, GNN is suitable for identifying the graph isomorphism, where the graph topologies and the vertex features of the compared graphs are similar. In some cases, the graph topologies and vertex features of a pair of binary code do not change much. For instance, the binary code change between binaries compiled under x86 and x64 is small. In this condition, GNN-based \BCSD approaches are effective. 
\begin{equation}
	\label{AGGREGATE}
	\vec{h}_i^{t+1} =AGGREGATE  \left(   \sum_{r \epsilon \mathcal{N}_i}\vec{h}_r^t  \right)
\end{equation}

\begin{equation}
	\label{Union}
	\vec{\mu}_i^{t+1} = Union \left( \vec{h}_i^{t+1} , \vec{\mu}_i^{t}) \right)
\end{equation}

\begin{equation}
	\label{READOUT}
        \vec{h}_G = READOUT \left(  \sum_{i \epsilon {G}}\vec{\mu}_i^{t+1}  \right)
\end{equation}

Unfortunately, due to the varieties of architectures and compiler options,
\emph{some logically similar binary code pieces appear to be dramatically different in the graph topology}. In this condition, the direct usage of GNN needs to overcome the challenges brought by significant binary change. 
Moreover, the $READOUT$ functions in GNNs are summation functions, which inevitability leads to an ``embedding collision'' problem---the summation of different node embeddings happens to be similar. For instance, when using Gemini to generate function embedding and performing function search, the embedding of query function $openssl\_set\_keygen\_ctx$, which contains 48 basic blocks, is very similar to the repository function $openssl\_crl2pkcs7\_main$ which includes 92 basic blocks. We checked and compared the neuron output in Gemini and found that these two functions' vertex embeddings are different. Their function embeddings are similar just because of the summation in $READOUT$. 
To further investigate the influence of ``embedding collision'', we manually analyze the FPs in the top 5 search results of 100 random function queries. We find that 95\% FPs are caused by the ``embedding collision''. The remaining 5\% FPs are TPs caused by function rename discussed in~\autoref{subsub:EvaluationacrossArchitectures}.
Thus, we believe that ``embedding collision'' is the primary limitation of the GNN-based \BCSD methods. {Notably, ``embedding collision'' uncovered in this paper is different from the GNN over-smoothing issue~\cite{chen2020measuring,cai2020note,rusch2023survey}, which causes two node embeddings with different labels presenting similar node embeddings (More comparison can be found in the artifacts).  }


\subsubsection{CNN}
\label{analysis-of-CNN}

Recent research~\cite{yu2020order,liu2018alphadiff} proposes to use CNNs on the adjacency matrices of the graph to capture the node ordering information of binary code. The intuition is that the node order changes of many cross-platform binary function pairs compiled from the same source code are small. Thus, the adjacency matrices of such function pairs should be similar. The features (node ordering) in the adjacency matrices can be recognized through convolution layers in a CNN model, which has an excellent performance when dealing with structured data such as image pixels by using different kernels, i.e., feature extractors~\cite{albawi2017understanding}. 
Unfortunately, though the CFG of a binary function can be transformed as an adjacency matrix just like the image pixels, there exist two primary characteristics when using  CNN models to extract the structural features of binary functions. 
(1) The number of basic blocks in binary functions is relatively small (90\% of the functions have less than 100 basic blocks).
(2) The connected relations among basic blocks are simple. 
Consequently, \textit{when searching repositories with a large number of functions, CNN-based approaches tend to report different binary functions with similar CFGs/adjacency matrices in the top search list.}  


\begin{figure*}[htbp]
    \begin{minipage}[t]{\textwidth}
        \centering
            \subfloat[ Gemini  threshold = 0.478]{
                \label{gemini-best-threshhold-for-testdataset} 
                \includegraphics[width=0.3\textwidth]{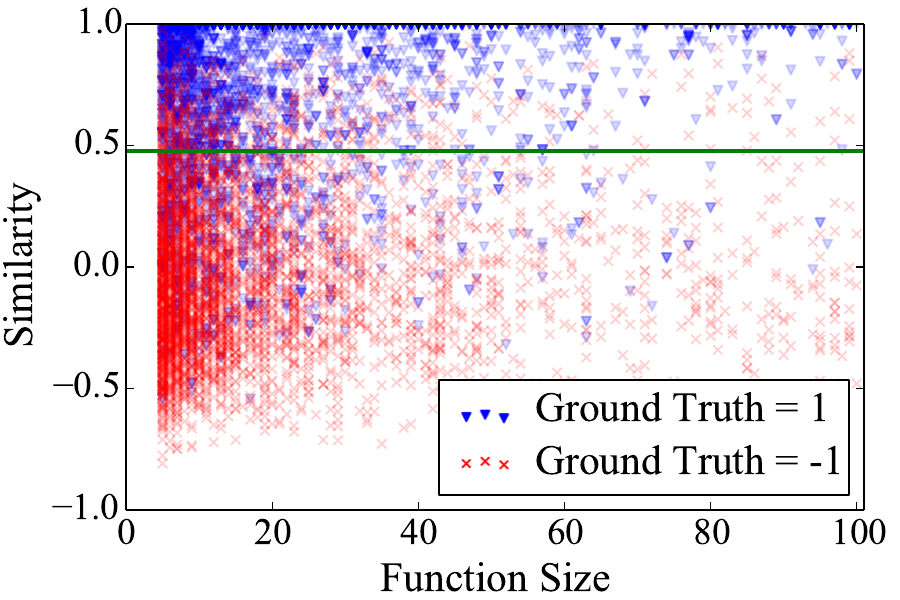}}
            \subfloat[ BinaryAI-skipt  threshold =0.562]{
                \label{ordermatters-best-threshhold-for-testdataset} 
                \includegraphics[width=0.3\textwidth]{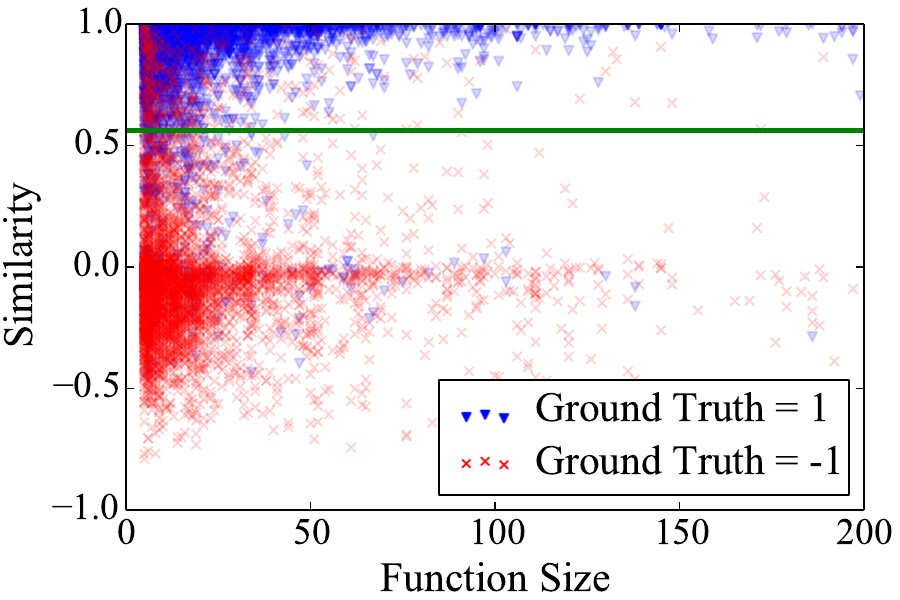}}
            \subfloat[ Focus-skip   threshold = 0.801]{
                \label{focus-best-threshhold-for-testdataset} 
                \includegraphics[width=0.3\textwidth]{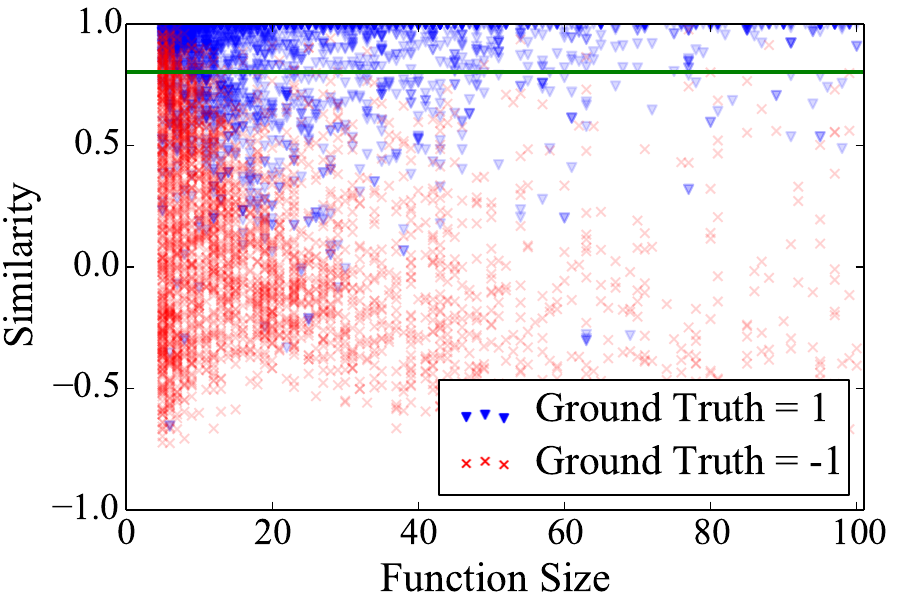}}
                
             \subfloat[Gemini function search ]{
                \label{gemini-best-threshhold-for-testdataset-1} 
                \includegraphics[width=0.3\textwidth]{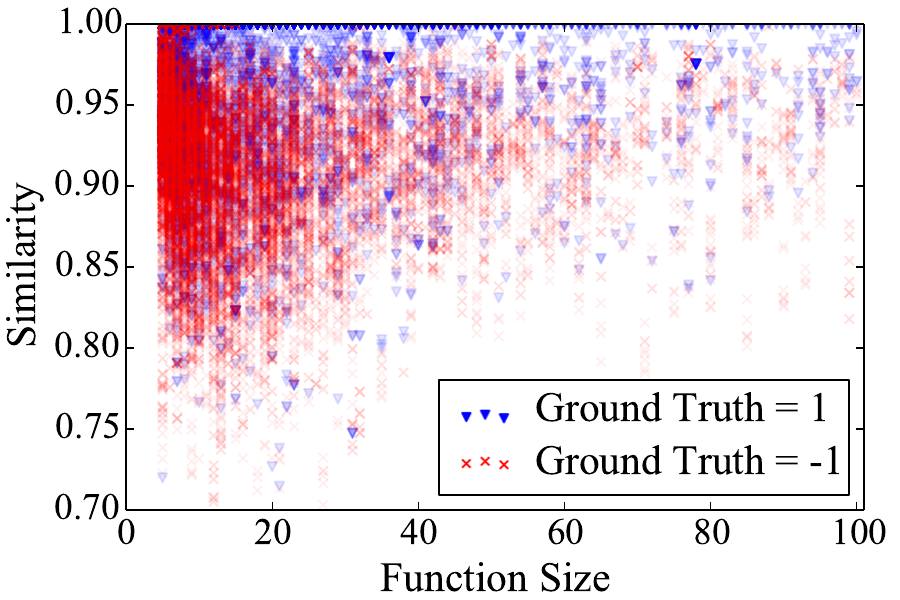}}
            \subfloat[BinaryAI-skipt function search ]{
                \label{ordermatters-best-threshhold-for-testdataset-1} 
                \includegraphics[width=0.3\textwidth]{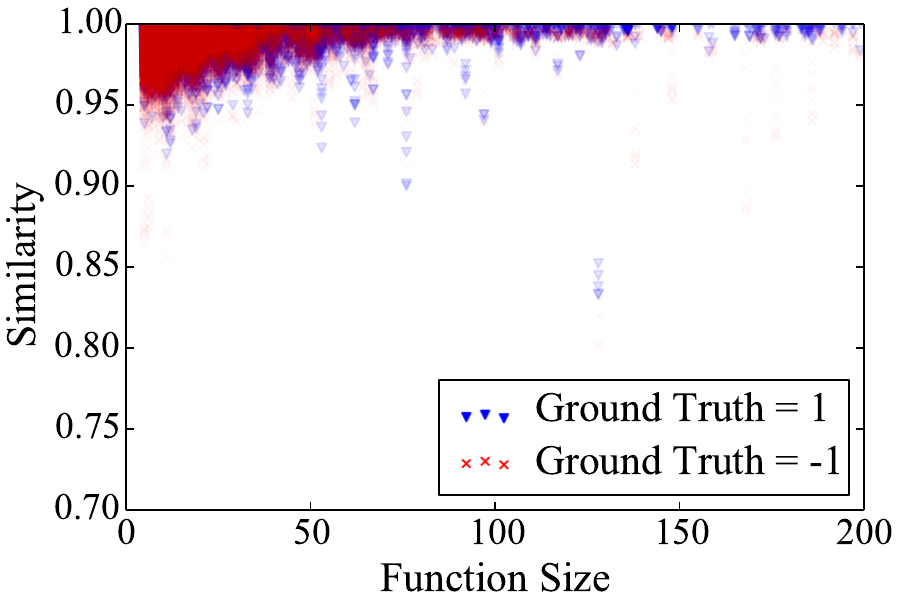}}
            \subfloat[Focus-skip function search ]{
                \label{focus-best-threshhold-for-testdataset-1} 
                \includegraphics[width=0.3\textwidth]{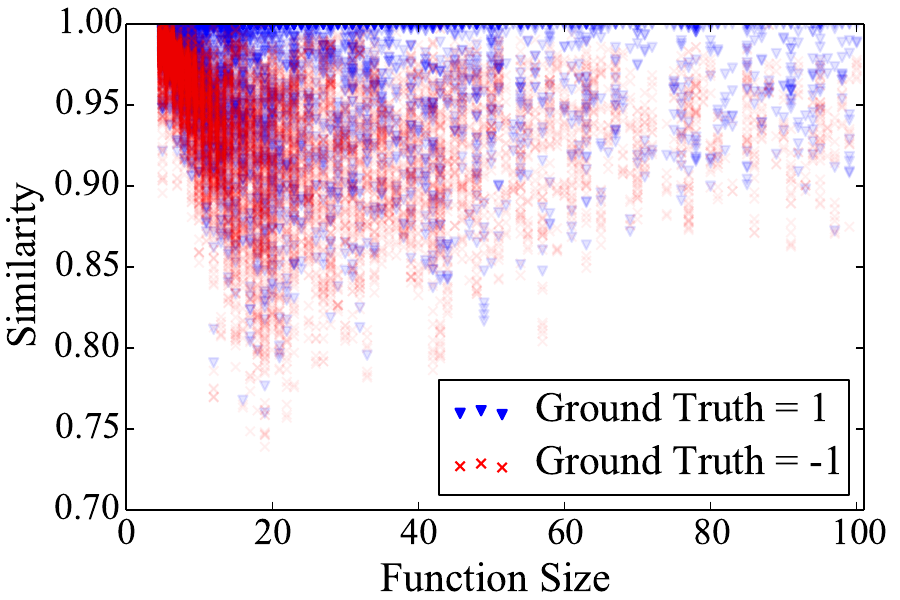}}
    \end{minipage}
        \caption{We perform AUC calculation on the test dataset.  Each point in (a), (b), and (c) presents the similarity score of each function pair in the test dataset. (d), (e) and (f) show the top-10 function search results in which the query functions are the first functions in the function pairs stored in the test dataset. Each point in (d), (e), and (f) present the similarity score between the query function and one of the search functions in the top-10 search result.}
        \label{fig:best-threshhold-for-testdataset}
\end{figure*}

\subsection{Evaluation Metrics}
\label{sub:evaluationmetrics}
In~\autoref{tab:BCSDs-comparision-SS-AUC}, many \BCSD approaches obtain high AUC values (larger than 0.98), indicating the performance of the corresponding embedding neural networks is good. However, when performing similar function detection, the precision and recall are not as good as the AUC values. In this section, we explore the calculation processes of AUC, precision, and recall to understand the performance gap between machine learning metrics and ranking metrics.

It is well known that the higher the AUC (near 1) is, the better the performance of the classification model is, which means that the model can choose a proper threshold to differentiate the TPs and TNs and minimize the FPs and FNs. Generally, AUC is obtained by evaluating a test dataset (randomly generated function pairs whose ground truth is labeled to be 1 if they are compiled from the same source code. Otherwise, the ground truth is -1). In the test dataset, the similarity score of two unrelated binary functions is usually low. The root reason is that \BCSD approaches are unlikely to determine two randomly selected binary functions compiled from different source codes to be similar since the occasionally selected functions are usually dissimilar in graph topology and function features. By contrast, the similarity score of two binary functions compiled from the same source code is usually higher than a certain threshold. Thus, \BCSD systems can easily find a threshold to classify similar function pairs and obtain a good AUC value. For instance, as shown in~\autoref{gemini-best-threshhold-for-testdataset}, ~\autoref{ordermatters-best-threshhold-for-testdataset}, and~\autoref{focus-best-threshhold-for-testdataset}, we record the similarity scores of function pairs in the test dataset of each \BCSD approach. We find that Gemini,  BinaryAI-skipt, and Focus-skip, can achieve high AUC values because models can obtain suitable thresholds (0.478, 0.562, and 0.801, respectively ) to maximize TPs and TNs.

However, in similar function detection, \BCSD usage is more similar to the recommendation process. A query function is compared with many other functions (100,000 in accuracy evaluation). In this condition, \BCSD approaches need to determine whether the top-K functions (which have high similarity scores) are compiled from the same source code as the query function. This task is more difficult than determining whether two randomly selected functions are compiled from the same source code. Consequently, for a \BCSD approach, a high AUC value can hardly demonstrate good search performance in function search. 
For instance, as shown in~\autoref{gemini-best-threshhold-for-testdataset-1}, ~\autoref{ordermatters-best-threshhold-for-testdataset-1}, and~\autoref{focus-best-threshhold-for-testdataset-1}, when performing function searches, it is difficult to differentiate TPs and TNs in the repositories because different binary code snippets also have high similarity scores when compared to the query function.


\section{Future Directions}
\label{s:futuredirection}
In the preprocessing phase, the extracted features can greatly influence what is to be learned by a neural network. High-quality binary features can absolutely benefit \BCSD neural networks to generate representative code embedding. Thus, we first discuss the promising direction of feature selection. In the code embedding process,  we conclude that the widely used GNN is more suitable for the \BCSD problem in ~\autoref{sbu:embeddingneuralnetworks}. However, the embedding collision problem significantly impedes the performance of GNN-based \BCSD approaches.
Thus, we next discuss potential mitigation strategies to alleviate the embedding collision problem.

    \textbf{Feature extraction by robust NLP models.} 
    Compared to manually extracted features obtained by experts, the features automatically extracted by robust NLP approaches contain more binary semantics. Particularly with the rapid development of NLP, many generative pre-trained transformers (GPT)~\cite{floridi2020gpt} that can follow complex instructions have been proposed to solve difficult problems with high accuracy. Similar to NLP, \BCSD also needs precise semantic understanding and extraction. Transferring the state-of-the-art large language models (LLM)~\cite{jiao2023chatgpt,tan2023evaluation,li2023starcoder} to the \BCSD problem is a promising future direction.

     \textbf{Embedding concatenation}. Embedding concatenation, which splices different types of embeddings, has been shown to be effective in structured prediction tasks~\cite{wang2020automated}. This technique may help alleviate the embedding collision problem because the concatenated embedding contains more binary semantics benefiting the differentiation of query functions and FPs (different binary functions that happen to be represented similarly to the query function due to the summation of basic block embeddings). For instance, we can concatenate the embedding of neural networks and constant strings embedding (likely unchanged features across various compiler options) as function embedding.
    %
     
    \textbf{Partial graph alignment.}  Despite optimization levels and architectures altering the structure and basic blocks of the graph representation of binaries, the graph representation of similar binary functions (such as CFG) is expected to be partially aligned since they are compiled from the same source code. For FPs caused by embedding collision, their graphs can not be aligned with the query function. Thus, intuitively,  graph alignment is a good way to identify FPs. In this paper, we perform a simple graph alignment---detecting identical basic block pairs (by comparing the basic block attributes obtained by each \BCSD method) in the search functions and query function to filter out FPs. As shown in ~\autoref{tab:BCSDs-graph-alignment}, this simple graph alignment significantly improves the precision of GNN-based \BCSD approaches, indicating that graph alignment is promising to mitigate the embedding collision problems.


\begin{table}
    \scriptsize
    \color{\tablecolor}
	\centering
	\caption{The change of precision and recall after performing graph alignment for GNN-based \BCSD approaches. NPrecision and NRecall present the new precision and recall after graph alignment.}
	
	\label{tab:BCSDs-graph-alignment}\vspace{-6pt}
	\setlength{\tabcolsep}{0.6mm} 
 	{
	\begin{tabular}{lcccc}
		\toprule
		\multirow{1}{*}{\textbf{\shortstack{Detector}}}   & \textbf{\shortstack{Precision@5} (\%)} &\textbf{\shortstack{Recall@5} (\%)}&\textbf{\shortstack{NPrecision@5} (\%)}
		 &\textbf{\shortstack{NRecall@5} (\%)}\\
		
		\midrule
            \rcolor Gemini-skip~\cite{fu2021focus}  & 41.85 & 25.42 & 54.39 & 25.38  \\ 
            VulSeeker-skip~\cite{fu2021focus}& 39.90& 23.87 & 49.82 & 23.50  \\ 
            \rcolor Focus-skip~\cite{fu2021focus} & {40.55} & {24.95} & 59.89 & 25.55  \\
            Gemini~\cite{xu2017neural}  & 44.32 & 27.10 & 56.99 & 26.90  \\   
            \rcolor VulSeeker~\cite{gao2018vulseeker}& 34.09& 21.32 & 50.16 & 22.34  \\ 
             Focus-mfe~\cite{fu2021focus} & {41.62} & {24.97} & 57.66 & 25.20  \\
           \rcolor MGMN~\cite{ling2021multilevel} & 29.99& 19.11 & 62.78 & 25.50   \\ 
           
		\bottomrule
	\end{tabular}}
\end{table}
\section{Related Work }
\label{s:discussion}

To investigate the \BCSD problem, recent work~\cite{haq2021survey} summarizes the characteristics and implementation details of  \BCSD approaches over the past 20 years. However, this measurement work does not involve a quantitative comparison of \BCSD approaches. Thus, this work fails to provide a straightforward and fair evaluation since most \BCSD approaches adopt different embedding strategies, evaluation methodologies, running environments, and/or benchmarks, making \BCSD approaches not directly comparable. 
As the first quantitative measurement in the \BCSD field, Marcelli et al.~\cite{marcellimachine} performed a systematic study to compare the effectiveness and efficiency of existing \BCSD approaches.
However, this measurement misses or lacks the exploration of multiple other important aspects, such as the in-depth understanding of the embedding neural networks and evaluation methodologies. In comparison (as shown in~\autoref{tab:Differences}), this work is novel in the following aspects. 

(1) New evaluation goals and challenges: Marcelli et al. overcome reproducibility issues and perform a fair comparison to reveal how different \BCSD approaches perform under different tasks (XO, XC, XA, etc.).
Our work overcomes new challenges, including understanding complex embedding processes and diverse evaluation methodologies (both are important for \BCSD development) to \textit{(a) shed light on the characteristics of mainstream embedding neural networks and the adopted evaluation metrics} and \textit{(b) investigate the root causes}. 
%
%
(2) New downstream applications: except for vulnerability search, we conduct the first license violation detection evaluation and identify the characteristics of different applications. For instance, compared to vulnerability detection, existing BinSD approaches can achieve better performance in license violation detection.   Moreover, we also summarize which \BCSD approaches are applicable to which downstream scenarios. 
(3) Novel findings: we present several findings to reveal the advantages and limitations of current embedding neural networks and evaluation methodologies and the corresponding root causes. For instance, despite GNN-based \BCSD approaches performing the best, their intrinsic limitation is the embedding collision problem. Moreover, evaluation methodologies such as repository construction can significantly influence the performance of \BCSD approaches. 

\begin{table}
    \scriptsize
    \color{\tablecolor}
	\centering
	\caption{Comparison to the related work.}
	
	\label{tab:Differences}\vspace{-6pt}
	\setlength{\tabcolsep}{0.6mm} 
 	{
	\begin{tabular}{llcc}
		\toprule
		\textbf{\shortstack{Type}} &\textbf{\shortstack{Task}} & \textbf{\shortstack{~\cite{marcellimachine}}} & \textbf{\shortstack{This paper}}\\
		
		\midrule
          \rcolor& The analysis of fuzzy hashing methods   & \checkmark & \ding{55} \\ 
           \rcolor \multirow{-2}{*} {Approach selection}  & The analysis of AI-powered \BCSD methods        &   \checkmark     & \checkmark\\
             & Feature extraction understanding & \checkmark& \ding{55}\\ 
              \multirow{-2}{*} {Skill understanding} & In-depth embedding network understanding & \ding{55}  & \checkmark \\
        \rcolor   Metric understanding  & In-depth metric understanding & \ding{55}  & \checkmark \\ 
            & The analysis of repository construction  & \ding{55}  & \checkmark \\  
             & The analysis of function rename  & \ding{55}  & \checkmark \\  
              & The analysis of unseen dataset & \ding{55}  & \checkmark \\  
             \multirow{-3}{*}{Evaluation} & Vulnerability detection   &   \checkmark     & \checkmark\\
            & License violation detection & \ding{55}  & \checkmark\\ 
            \rcolor   Future work  & The validation of  {promising directions} & \ding{55}  & \checkmark \\ 
		\bottomrule
	\end{tabular}}
\end{table}

\section{Conclusion}
\label{s:conclusion}

In this paper, we perform a systematic measurement of the state-of-the-art AI-powered \BCSD approaches.
First, we conduct an extensive evaluation to fairly and quantitatively understand the true capability of \BCSD tools in similar function detection and two main downstream applications. Moreover, we perform an in-depth analysis of binary embedding neural networks and evaluation methodologies to shed light on their pros and cons. 
\LPY{The findings and implications of our study are expected to boost a better understanding of the status-quo of AI-powered \BCSD.} 
Finally, we provide and validate several promising future directions for advancing \BCSD. We hope the open-source of our datasets, benchmarks, and implementation can facilitate the development of \BCSD.

\clearpage
\bibliographystyle{ACM-Reference-Format}
\bibliography{p}


\end{document}